\documentclass[twocolumn, dvipsnames]{aastex63}

\usepackage{longtable}
\usepackage{upgreek}
\usepackage{hyperref}
\revised{\today}

\submitjournal{ApJ}

\shorttitle{Sample article}
\shortauthors{Prince et al.}
\graphicspath{{./}{figures/}}

\newenvironment{myfont}{\fontfamily{lmtt}\selectfont}{\par}
\DeclareTextFontCommand{\textmyfont}{\myfont}
\def\kms{\,km\,s$^{-1}$}

\begin{document}

\title{Viewing angle observations and effects of evolution with redshift, black hole mass, and Eddington ratio in quasar based cosmology}


\author[0000-0002-1173-7310]{Raj Prince}
\email{raj@cft.edu.pl}
\affiliation{Center for Theoretical Physics, Polish Academy of Sciences, Al.Lotnikow 32/46, 02-668 Warsaw, Poland}
\author[0000-0002-2005-9136]{Krzysztof Hryniewicz}
\affiliation{National Centre for Nuclear Research, ul. Pasteura 7, 02-093 Warsaw, Poland}
\author[0000-0002-5854-7426]{Swayamtrupta Panda}
\affiliation{Center for Theoretical Physics, Polish Academy of Sciences, Al.Lotnikow 32/46, 02-668 Warsaw, Poland}
\affiliation{Nicolaus Copernicus Astronomical Center, Polish Academy of Sciences, ul. Bartycka 18, 00-716 Warsaw, Poland}
\author[0000-0001-5848-4333]{Bo\.zena Czerny}
\affiliation{Center for Theoretical Physics, Polish Academy of Sciences, Al.Lotnikow 32/46, 02-668 Warsaw, Poland}
\author[0000-0003-3358-0665]{Agnieszka Pollo}
\affiliation{National Centre for Nuclear Research, ul. Pasteura 7, 02-093 Warsaw, Poland}

\begin{abstract}
This study is focused on the observational measurement of the viewing angle of individual quasars by modeling the broadband quasar spectrum ranging from the infra-red (IR) to soft X-ray band. Sources are selected from various published catalogs, and their broadband quasi-simultaneous spectral data points were collected and used to model. We started with COSMOS sample of type-1 sources which have broadband photometric points. To include more data points we cross-matched the COSMOS with the SDSS DR14 quasar catalog, and eventually we find 90 sources which have broadband data ranging from IR to soft X-ray.
The broadband spectral energy distribution (SED) modeling is done in \texttt{Xspec} by using the \texttt{optxagnf} and the \texttt{SKIRTOR} models for the X-ray, UV, Optical, and IR regimes for each source. 
The whole sample is divided into four bins with respect to redshift, black hole (BH) mass, and Eddington ratio with an equal number of sources in each bin. The viewing angle is estimated in each bin, and its evolution with respect to redshift, BH mass, and Eddington ratio is examined. In result, we did not find any significant evolution of viewing angle with those parameters
within the 95$\%$ confidence interval.
We conclude that the use of quasars in cosmology to determine the expansion rate of the universe is therefore justified, and biases are not expected.
\end{abstract}


\section{Introduction} \label{sec:intro}

 Recent detection of Hubble tension between the early universe \citep[]{Planck-Collaboration-2018-VI} and local universe - SNIa \citep[e.g.][]{riess2019,Riess2019NatRP,riess2020}, T-RGB \citep{freedman2019}, Masers \citep{Pesce_2020_MegaMaser} measurements has posed many serious questions on the standard cosmological model. Though the tension observed from the independent method is not claimed with 100 $\%$ clarification because it is also believed that the systematic error in each independent method can also lead to a tension in H$_0$.
In general, the issue has become a key question for present-day astronomy and physics. However, before any conclusion about the need for new physics is drawn, improvements in the local measurements are clearly needed. In order to 
provide robust conclusions, we need to have many independent individual measurements of the Hubble constant, and we have to carefully discuss the biases that might be present in each method.  Many others measurements are already at work based, for example, on Baryon Acoustic Oscillations (BAO; \citealt{bassett_hlozek_2010}), gravitational lensing (\citealt{10.1093/mnras/stz3094}), gamma-ray bursts (\citealt{Schaefer_2003}), and also active galactic nuclei (AGN, \citealt{2019NatAs...3..272R}) as a cosmological tool, and recently also on the detection of gravitational waves (\citealt{2019BAAS...51c.310P}). 

Studying cosmological parameters at the local universe using the so-called 'standard (or standardizable) candle' approach requires to have a way to measure the source's absolute luminosity and good coverage of redshift. Quasars or, more generally,  AGN fulfill the criteria and are suitable 'standardizable candle' to do cosmology utilizing wider coverage of redshift. Though their luminosities vary by several order of magnitude, there are ways to determine the absolute luminosity of each quasar individually, on the basis of their spectral energy distribution \citep{risaliti2015}, continuum time delays \citep{collier1999,cackett2007}, and emission line time delays \citep{watson2011,haas2011,czerny2013}. 
However, there is an uncertainty inherent to observations of quasars and we need to take care of that while doing cosmology. Quasars are not isotropic sources, and their observed luminosity depends on the viewing angle, which is highlighted in the diversity in their emission line profiles (a considerable part of the incoming radiation gets intercepted by intervening medium, such as the obscuring torus, which leads to the sub-division of quasars where broad emission lines are seen in their spectra, i.e., Type-1 sources, or not, i.e. Type-2 sources).

In the standard scenario, authors use a single average viewing angle for all the sources, which span in a wider range of redshift, mass, and Eddington ratios. In our previous study \citep{prince2021}, we have discussed how the viewing angle of an individual quasar can affect the measurement of the Hubble constant or any other cosmological parameters. Our results show that the viewing angle effect is important when one is trying to constrain the cosmological parameters using quasars. Dispersion in the viewing angles can simply cause dispersion in measurements, but any systematic change of the viewing angle with redshift can significantly affect the measurements of the cosmological parameters.

In \citet{prince2021}, we found relations between the possible systematic difference in the viewing angle of the low and the high redshift quasars, but the estimates were based on toy model considering the two quasar samples by \citet{gu2013}.

In the present paper, we estimate the viewing angles of individual quasars by modeling the broadband spectral energy distribution (SED) for a sample of quasars. There are very few sources that have multi-wavelength observations in X-ray, UV, optical, and IR, one that is important for the SED modeling. In the absence of such multi-wavelength data, it is a challenge to constrain the model parameters of individual sources in practice. Hence, we estimate the cumulative parameters by dividing the considered source sample into four different bins with respect to redshift, BH mass, and Eddington ratios. For the quasar-based cosmological model, the main parameter here is the viewing angle of the quasar. If we detect the redshift evolution of the viewing angle, then the measurement done in the past or planned in the future adopting the same average viewing angle for all redshift sources are incorrect, and they need to be modified. However, if we do not see any evolution within the errorbar of the measurement, we can say that cosmology is safe with the quasar.

\section{SAMPLE SELECTION}

To test the hypothesis, we prepare a sample of AGNs with photometric data available across a wide range of wavelengths (NIR to X-ray). We start with the \citet{marchesi+16} catalogue that consists of X-ray sources with the optical/NIR counterparts (i, K, and 3.6$\mu$m) for the COSMOS sample\footnote{\href{http://vizier.u-strasbg.fr/viz-bin/VizieR-3?-source=J/ApJ/817/34/catalog}{http://vizier.u-strasbg.fr/viz-bin/VizieR-3?-source=J/ApJ/817/34/catalog}}. The COSMOS sample is an X-ray selected ($\sim$ 3$\sigma$, 2-10 keV band) sample obtained from the combination of the 1.8Ms C-COSMOS survey \citep{elvis+09} with 2.8Ms observations from the COSMOS Legacy survey performed by \textit{Chandra} ACIS-I \citep{civano+16}. The \textit{full} COSMOS sample contains 4,016 X-ray point sources (2,273 from the COSMOS Legacy survey and 1,743 from the C-COSMOS survey). Of these 4,016 sources, 2,196 sources have reliable spectroscopic redshifts. These spectroscopic redshifts are obtained by cross-matching the sources with optical counterparts using a master spectroscopic catalog containing $\simeq$ 80,000 sources \citep[see][for more details]{marchesi+16}. This value-added catalog contains additional spectroscopic identification flags, which are used to filter for sources that are classified as Type-1 or Type-2 AGNs. Sources are classified as "BLAGN" or Type-1 if they show evidence of at least one broad (i.e., with FWHM $>$ 2000 \kms) line in their spectra. Sources with only narrow emission lines are categorized as "non-BLAGNs". No further separation is made between star-forming galaxies and Type-2 AGNs in the COSMOS sample. Combining the information of reliable spectroscopic redshifts and spectral type information, we have 638 Type-1 and 1,070 Type-2 sources.

To obtain additional optical photometric measurements, we cross-matched the Type-1 and Type-2 sources from the COSMOS sample with the SDSS DR14 quasar catalog\footnote{\href{http://vizier.u-strasbg.fr/viz-bin/VizieR?-source=VII/286}{http://vizier.u-strasbg.fr/viz-bin/VizieR?-source=VII/286}} \citep{paris+18}. In  addition  to  the  SDSS ugriz bands, the DR14 catalog provides multi-wavelength  matching  of  DR14  quasars  to several surveys: the FIRST radio survey \citep{becker+95}, the Galaxy Evolution Explorer \citep[GALEX,][]{martin+05} survey in the UV, the Two Micron All Sky Survey \citep[2MASS,][]{cutri+03,skrutskie+06}, the UKIRT Infrared Deep Sky Survey \citep[UKIDSS,][]{lawrence+07}, the Wide-Field Infrared Survey \citep[WISE,][]{wright+10}, the ROSAT All-Sky Survey \citep[RASS,][]{voges+99,voges+00}, and the seventh data release of the Third XMM-Newton Serendipitous Source Catalog \citep{rosen+16}. The result of the cross-match with the DR14 catalog gave us a rather limited number of sources -- 90 Type-1 and two Type-2 sources\footnote{The two Type-2 sources are actually "BROADLINE" type QSO according to their SDSS-BOSS spectra, but we do not include them in our study.}.

\section{Broadband SED Modeling}
\label{sec3}
As discussed in section 2, we have collected the sources from various catalogs with broadband observations ranging from IR to X-rays. We aim to determine the individual source parameters for all the sources in the sample by performing the broadband spectral energy distribution (SED) modeling. The broadband SED modeling is performed in \texttt{Xspec} \citep{xspec_1996}, and hence the data are prepared accordingly. The data are used to create the spectrum and the response files for the \texttt{Xspec}, using the tool \texttt{ftflx2xsp}. Finally, the spectrum files and the response files are loaded in \texttt{Xspec} for modeling.  
To model the optical, UV, and X-ray part of the spectrum, we have chosen the inbuilt \texttt{Xspec} model, e.g., \texttt{optxagnf} 
which is developed by Chris Done and her collaborators. The detailed description of the model along with the parameters is given in \citet{Chris_2012}. The model has eleven parameters, and among them, most of the parameters are fixed to the standard value. 
The parameters in the \texttt{optxagn}  are following,
mass of black hole (M$_{BH}$), Eddington ratio (L/L$_{Edd}$), luminosity distance (D$_L$) and the redshift ($z$) of the source, dimensionless black hole spin (a$_{star}$), coronal radius (r$_{cor}$), outer radius of the accretion disk (r$_{out}$), electron temperature (kT$_e$) and the optical depth ($t$) of the soft-Comptonization component, spectral index ($\Gamma$) of the hard-Comptonization component, and the fraction of the power (f$_{pl}$) in coronal radius emitted by hard-Comptonization component. Among them mass, distance, and redshift are fixed from the catalog value (see Table \ref{tab:my-table}) and along with that other parameters such as kT$_e$=0.2, a$_{star}$=0, log(r$_{out}$)=5 are fixed. The normalization is fixed at 1 which is suggested by the model itself assuming disk inclination 60 degrees. 
Initially, the parameters such as black hole mass and Eddington ratio have been kept free for the modeling. But while modeling the data we realized that in a few sources the number of total free parameters exceed the number of data points and hence to overcome this situation we decided to fix the mass of all the sources to the catalog value (Table \ref{tab:my-table}). After fixing the mass we are left with the total number of free parameters equal to 12 and the total number of photometric data points in SEDs are 13. There are 10 sources where the photometrc data points are less than the number of parameters and hence only 80 sources out of 90 are fitted.

To model the reprocessing of the disk emission by the torus, we have used the \texttt{SKIRTOR}\footnote{ \href{https://sites.google.com/site/skirtorus/sed-library}{SKIRTOR}} model, which models the torus as a two-phase medium. It considers a large number of high-density clouds embedded in the smooth low-density medium. The concept of a two-phase medium for torus is supported by many hydrodynamical simulations performed by \citet{Wada_2009} and \citet{Wada_2012}. Along with that, many observational studies also suggest the possibility of a two-phase medium (\citealt{Assef_2013}).
The detailed description of the \texttt{SKIRTOR} model is given in \citet{Stalevski_2012} and \citet{Stalevski_2016}. This model is not part of the \texttt{Xspec} and hence we have to prepare the model to load into \texttt{Xspec}. 
The SKIRTOR library is then included into \texttt{Xspec}.
\texttt{SKIRTOR} model has a total of seven parameters, and the most important parameters for our study are (1) the optical depth along the observer line of sight, (2) the opening angle of the torus, and (3) the viewing angle. The viewing angle is the angle between the observer's line of sight and the BH symmetry axis. The geometrical description of the torus is shown in Figure 1 of \citet{Stalevski_2012}.
The inclination of the disk or the viewing angle in these quasars can not be estimated directly from the Fe K-alpha line in the X-ray spectrum because of two following reason: first, not all of the sources have good quality X-ray spectra, and secondly, at higher redshift Fe K-alpha line is not properly constrained in X-ray spectra since sources are not very bright. Hence, we decided to measure the viewing angle with the help of the torus geometry.
The torus is assumed to have a gradient density distribution of matter along the polar as well as in the radial direction. 

\begin{figure}
\vspace{0.2in}
    \centering
    \includegraphics[scale=0.35, angle=-90]{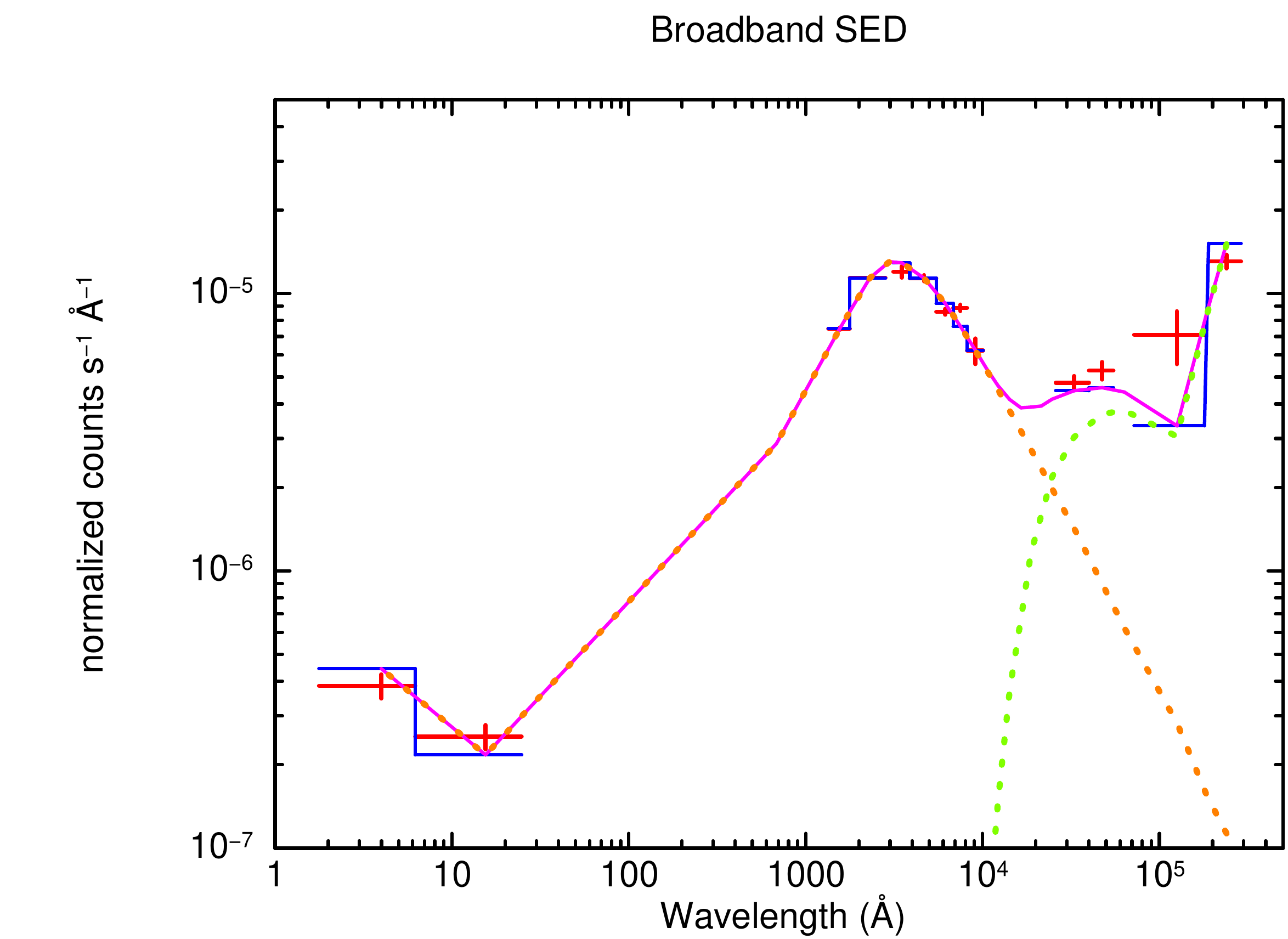}
    \caption{Broadband SED ranging from X-ray to IR, modeled with \texttt{optxagnf} and \texttt{SKIRTOR} models. Red data points are the spectral points, yellow, and green dotted lines are \texttt{optxagnf} and IR part of the \texttt{SKIRTOR} model, respectively, and the blue and red color bold lines are the total model fitting just the data in chunks and the overall data, respectively.}
    \label{fig:sed}
\end{figure}

The \texttt{SKIRTOR} model assumes the accretion disk as a point source which is the primary source of emission with an an-isotropic emission pattern. The shape of the primary source is a power-law. As suggested in the \texttt{SKIRTOR} tutorials, if the interested source has a different SED shape, one can always replace the power-law with the best fit SED shape. In our case, we have replaced the power-law with the \texttt{optxagnf} model in order to explain the primary source emission, which is defined as a multiplicative model \texttt{m1} in our modeling. Now to suppress the primary emission component from the \texttt{SKIRTOR}, we have prepared a multiplicative model \texttt{m2} which is basically the ratio of transmitted emission to the incident emission defined in \texttt{SKIRTOR} model. A separate total dust emission model \texttt{m3} is prepared as an additive model in order to model the infrared (IR) emission in broadband SED.

To model the broadband SED, we have used following models-
\begin{enumerate}
\item { \texttt{redden}:} to account for the interstellar extinction in UV
\item { \texttt{optxagnf} (\texttt{m1}):} to account for optical-UV, and X-ray emission from the primary source
\item {\texttt{SKIRTOR} model (\texttt{m2}):} ratio of scattered to incident emission 
\item {\texttt{SKIRTOR} model (\texttt{m3}):} total dust emission.
\end{enumerate}
The combination of all the above models is finally used to fit the broadband SED. The total model has the form:  (\texttt{redden}*\texttt{m1}*\texttt{m2}) + \texttt{m3}. The first part of the expression takes care of the optical-UV and X-ray emission, and the second part models the IR emission. The modeling parameters for the \texttt{m2} and \texttt{m3} are common and hence are kept equal to each other. 

During the whole modeling procedure the mass of the objects are fixed from Table \ref{tab:my-table}, and the Eddington ratio is free in \texttt{optxagnf} model. The outer radius of the disk at 10$^5$R$_{g}$, and the electron temperature of the soft Comptonisation component is also fixed at 0.2 keV in \texttt{optxagnf} model. In the  \texttt{SKIRTOR} model (\texttt{m2} \& \texttt{m3})
all the parameters are free except redshift. 
A successful broadband SED modeling is performed, where the chi-square is minimized, for all the sources except 10 sources that do not have IR data 
in our sample. An exemplary SED model fit is shown
in Figure \ref{fig:sed}. Important parameters derived from the modeling are tabulated in the last three columns of Table \ref{tab:my-table}. The optical depth is estimated along the line of sight \citep[see Equation 1 in][]{prince2021}. 
As it can be seen in Table \ref{tab:my-table}, there are few sources that have very low optical depth ($\leq 2 \times 10^{-2}$) that can be considered as top-view sources, and the sources with high optical depths can be considered as highly inclined. We have also examined the absorption of central emission in the dusty torus for these two extreme situations. As expected, we see a small change in the optical/UV emission due to absorption in a highly inclined source (optical depth ($\tau$) = 4.430; Table \ref{tab:my-table}) while no absorption is seen in top-view sources ($\tau$ $\leq 2 \times10^{-2}$). 

In a few sources where number of data points are more, we have considered mass also as a free parameters. In result, we noticed that the fitted mass is lower than the mass mentioned in Table \ref{tab:my-table}.
The best fit values of the parameters and the corresponding chi-squared values are collected for further interpretation. The parameters we are interested in here are the viewing angle, optical depth, and the opening angle of the torus. All these parameters are part of \texttt{SKIRTOR} model. We have collected the fitted value of viewing angle from the model for all the sources. 
While doing the modeling, we noticed that most of the time viewing angle comes above 60 degrees (Table \ref{tab:my-table}). We ran the \texttt{steppar} procedure in \texttt{Xspec} for the viewing angle to have the $\chi^2$ distribution of the viewing angle. In this process, we notice that all the values below 60 degrees are also possible for viewing angle, and we did not find a proper single minimum but rather flat distribution at a lower viewing angle. This trend was seen in most of the sources, but not in all. As an experiment, we also included the far-IR data for a few sources from the Herschel Telescope and did the modeling. The result we got is very much expected, the viewing angle does not change after including the more far-IR data, which makes us to believe that the far-IR data are not sensitive to the viewing angle since the torus becomes transparent at those wavelengths. Only accurate spectroscopy sensitive to features could help, but that might be quite sensitive to model assumptions also.

We have collected the $\chi^2$ distribution of the whole allowed range of viewing angle (i.e. 0-80 degree). Running \texttt{steppar} for the viewing angle also changes the other parameters in the background, and we have also read the values of the optical depth and opening angle from those background parameters that vary with the viewing angle. During the modeling, we have realized that the opening angle of the torus can not go below 10$^\circ$, which is the limit put by the \texttt{SKIRTOR} model. In our results, in many sources, we encountered this lower limit of the opening angle (see Table \ref{tab:my-table}).

So to constrain the upper and lower limit of the viewing angle, we have estimated a Likelihood function (LF) for each source, which gives a probability distribution of viewing angle.
The Likelihood function is approximated as LF $\sim$ exp(-$\chi^2$). The shape of the probability distribution for each source is different and highly depends on the best fit value of the parameters. 
Since it was not reliable to estimate the individual viewing angle because of the large allowed range we have divided the total sample in four different bin, each bins consist of 20 sources. The results are discussed in the next section.

\section{Results}
\label{sect:results}

\begin{figure*}
    \centering
    \includegraphics[scale=0.26]{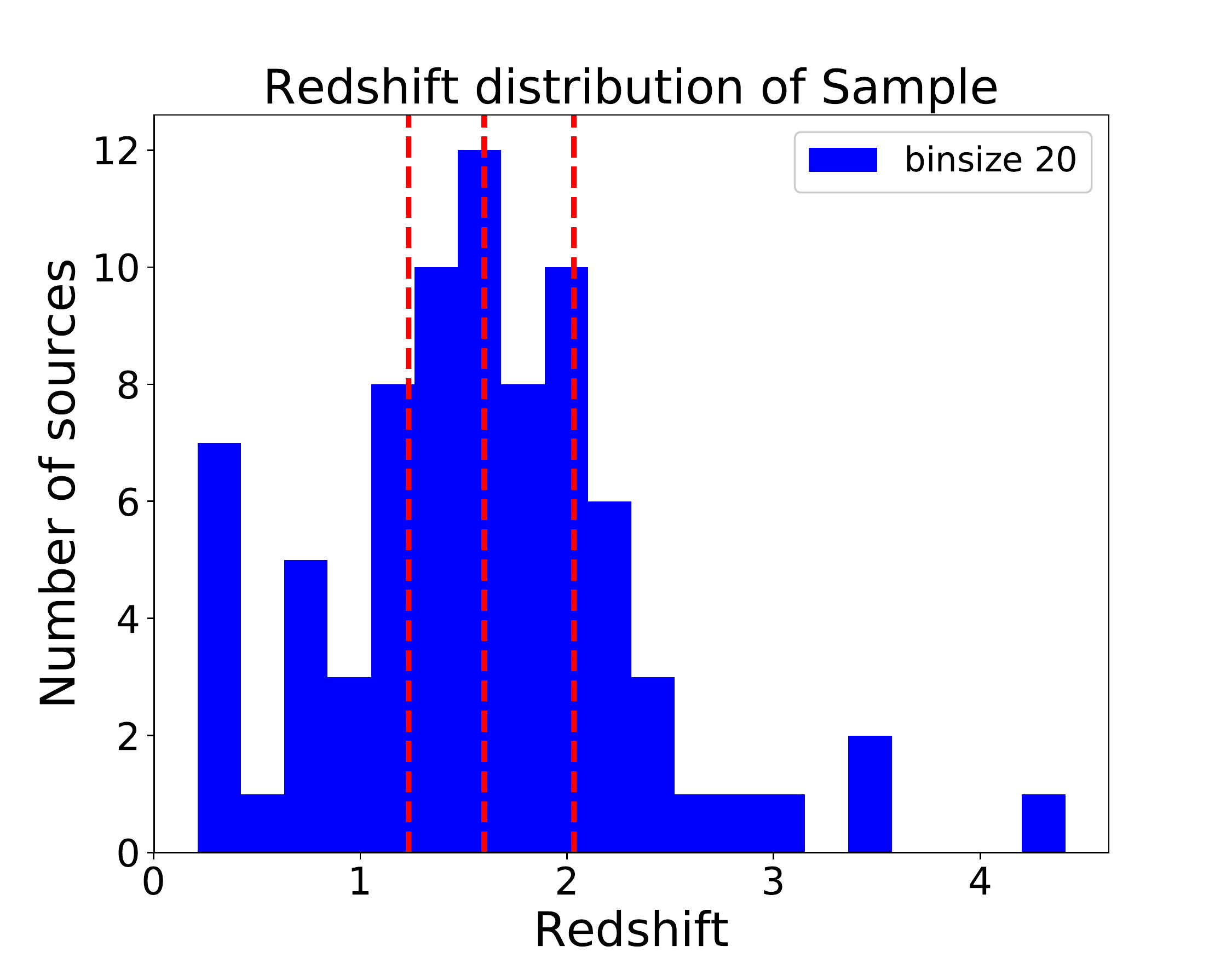}
    \includegraphics[scale=0.26]{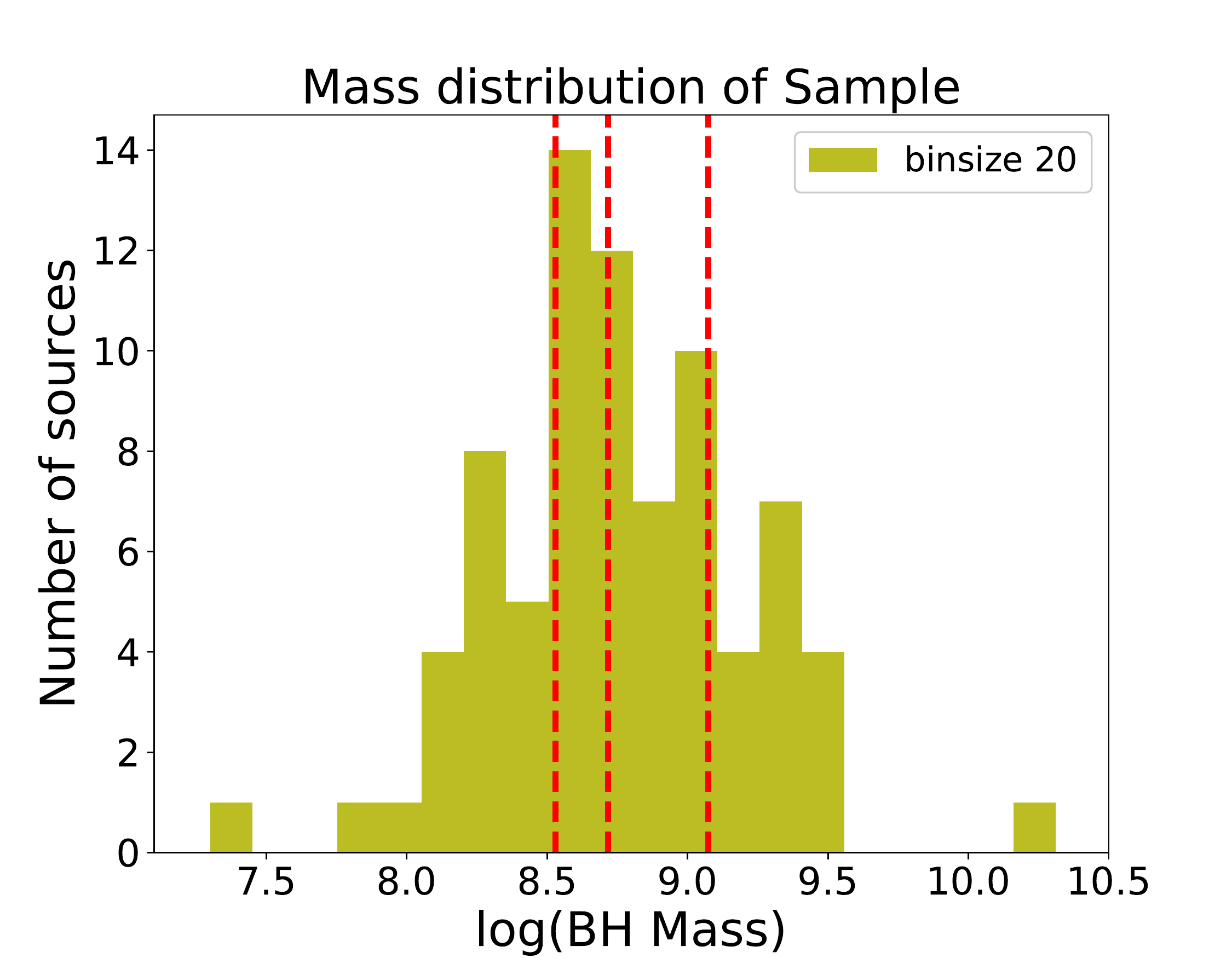}
    \includegraphics[scale=0.26]{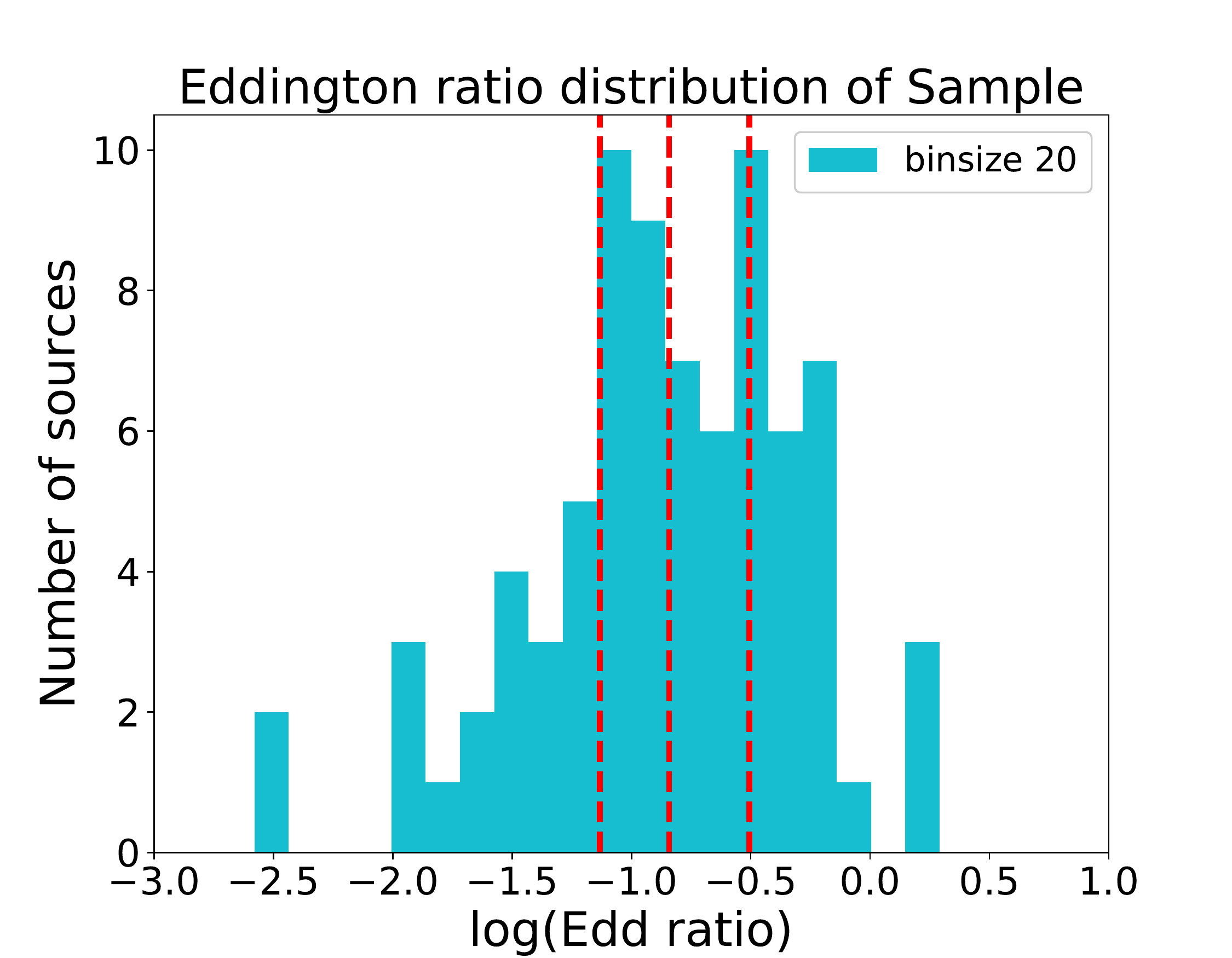}
    \caption{Distribution of our entire sample with respect to redshift, mass, and Eddington ratio are shown from left to right for equal number of bins.}
    \label{fig:distribution}
\end{figure*}

\begin{figure*}
    \centering
    \includegraphics[scale=0.29]{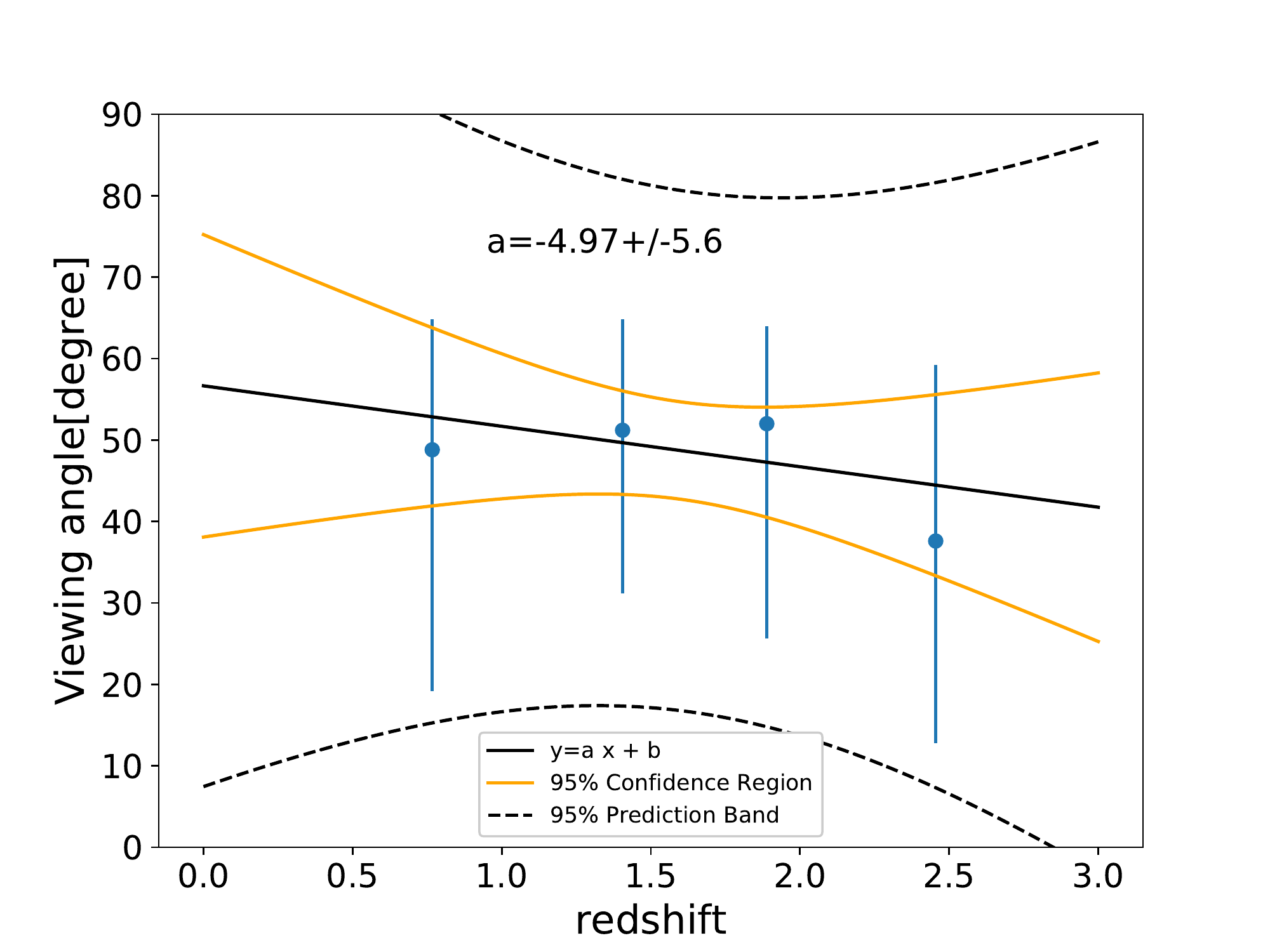}
    \includegraphics[scale=0.29]{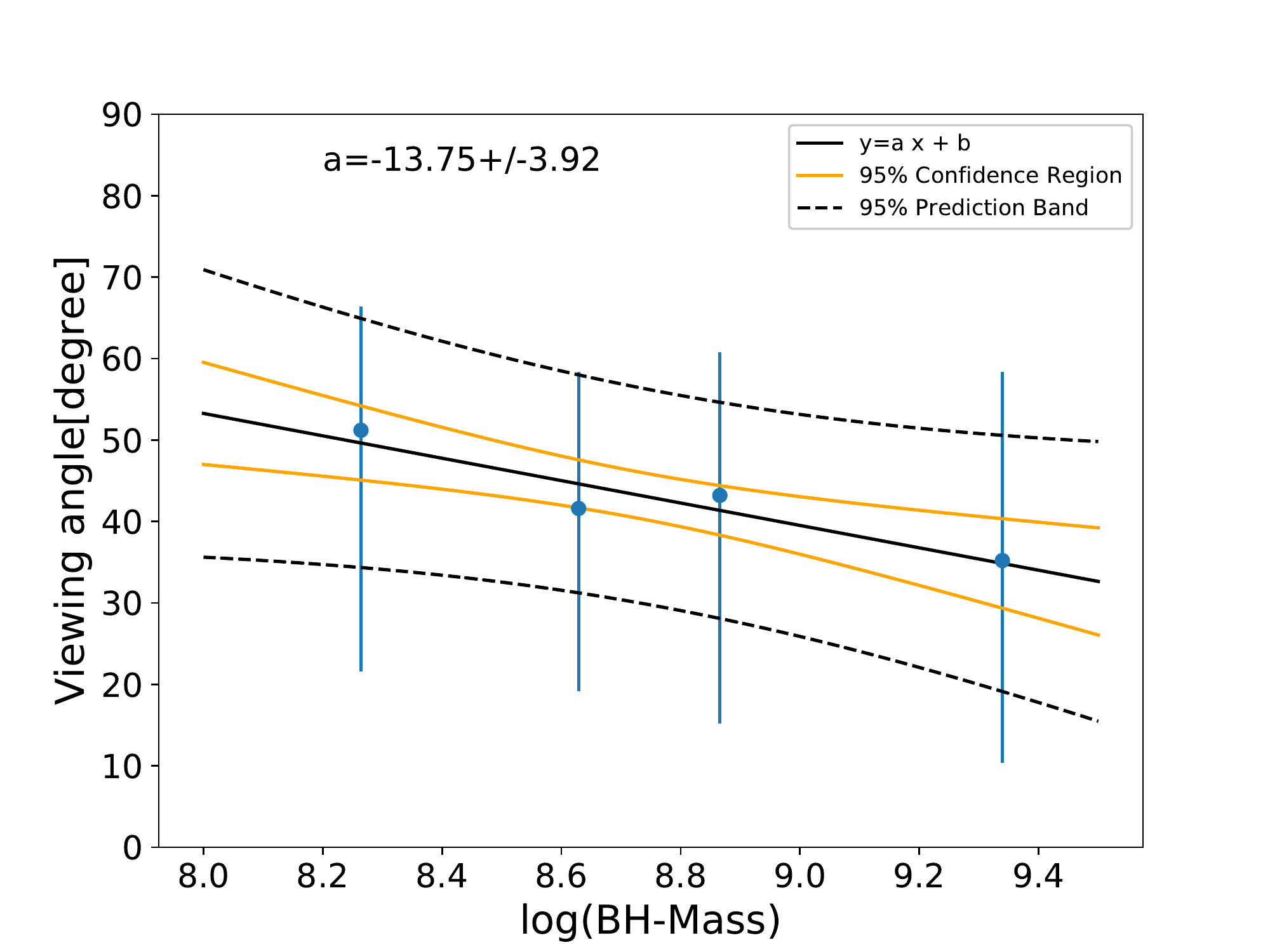}
    \includegraphics[scale=0.29]{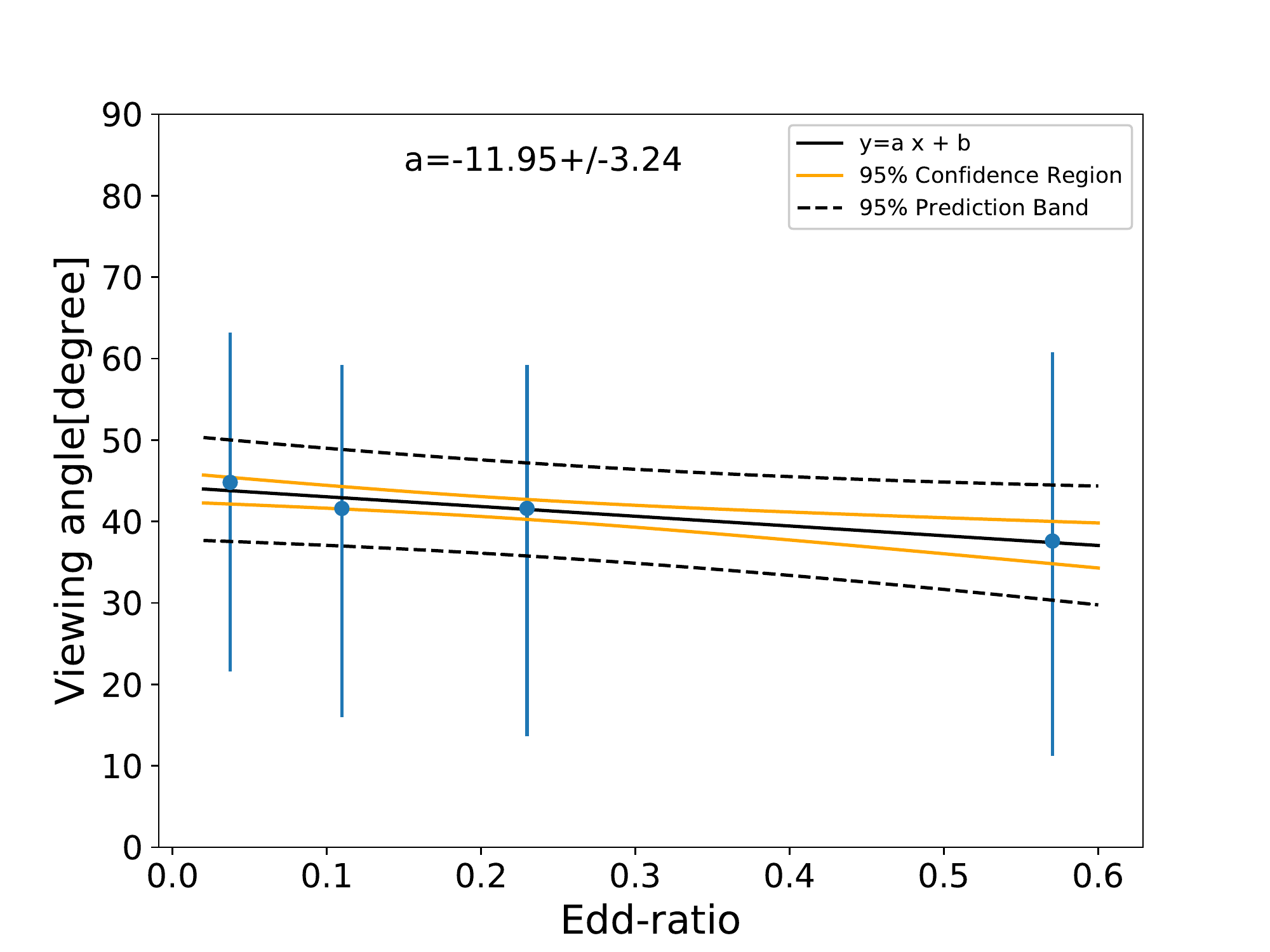}
    \caption{Viewing angle dependence on redshift. Viewing angles are estimated at median value of each redshift bins. The significance of the regression is above 90$\%$.}
    \label{fig:median}
\end{figure*}

We modeled the broadband SED of 80 active galaxies using a realistic model. We give the results for the key parameters for the individual sources in Table~\ref{tab:my-table}. The errors of the individual measurements are large because of the model complexity and a large number of parameters (in this case 12 parameters). The torus opening angle measured from the equatorial plane is found to be 10 degrees for several sources since this was the lower limit in the model. The model parameters are considerably degenerate if the dusty material is not located along the line of sight, since in that case, the torus opening angle and torus optical depth measured along the equatorial plane are interdependent. However, our goal was not in the precise determination of all parameters in each source but in tracing the global trends in the sample.

\subsection{Trends with redshift}

Since the main objective was to examine the viewing angle evolution with respect to redshift, we have divided our entire sample into four redshift bins with an equal number of sources in each bin also marked by red dotted vertical lines. The four redshift bins are 0.212--1.234, 1.2348--1.6003, 1.6079--2.0336, and 2.0541--4.412 with median value of each redshift bins 0.767, 1.405, 1.889, 2.455, respectively. The distribution of sources with respect to redshift is shown in Figure \ref{fig:distribution}. 
The number of bins was optimized to have the broad redshift range (better to have more bins) and the small statistical errors in each bin (fewer bins is better). The choice of four bins gives 20 sources per bin.  

The cumulative probability distribution is estimated for each bin by multiplying all the probability distributions of individual sources within a single bin. The distribution of normalized cumulative probability for the viewing angle, black hole mass, and Eddington ratio are shown in {\it Appendix A} (Figure \ref{fig:comulative_prob}) and for exemplary purpose, we also show normalized probability distribution of a single source in Figure \ref{fig:normprob}. Considering the 50$\%$ part of the cumulative distribution would give the location where the distribution peaks. The distribution is integrated for a total 50$\%$ contribution to get the best solution of viewing angle. Further error estimation is done by considering 1$\sigma$ (68.27$\%$) error within the limit, which covers both sides of the distribution. For the positive error, the integration is done from the positive side of the distribution for the one side contribution of 15.86$\%$. Similarly, the negative error is estimated from the other side of the distribution for the same contribution of 15.86$\%$. Following the above procedure, the best solution found for the viewing angle in each bin, and their corresponding positive and negative error are shown in Table \ref{tab:result}. The redshift dependent viewing angle is shown in Figure \ref{fig:median} and fitted with a straight line to examine the evolution. The slope of the fitted line is -4.31$\pm$5.05, which nicely covers the range of viewing angle from 37.6 to 52.0 within the errorbars. The viewing angle range obtained between 37.6 to 52.0 degree is mostly consistent with the type-1 quasars, which was also the sample we had chosen.
A similar procedure is followed for the mass and Eddington ratio. The best viewing angle estimated in each mass bin and Eddington ratio bins are also shown in Figure \ref{fig:median}, and the best values with errorbars are mentioned in Table \ref{tab:result}.
From Table \ref{tab:result}, it is clear that in the first three redshift bins, the viewing angle is found to be very close to each other. However, the redshift bin four has a viewing angle of 38.4 which is far from the first three bins. One of the reasons could be the large range of redshift considered in the fourth bin (2.0541--4.412). 

\subsection{Trends with BH mass and Eddington ratio}

Similarly, we have also divided the whole sample of sources into four bins of mass and Eddington ratio to examine the viewing angle evolution with respect to mass and Eddington ratio independently. The BH mass and the Eddington ratio taken here are the catalog values from Table \ref{tab:my-table} not the fitted value.
The median mass in each mass bin is 8.264, 8.629, 8.865, and 9.339 (in logarithmic units), and similarly, the median Eddington ratio in each bin is 0.0373, 0.1097, 0.2297, and 0.5703, respectively. Histograms of redshift, mass, and Eddington ratio distributions are shown in Figure \ref{fig:distribution}.

In the case of mass dependence of viewing angle, we found that the values are less scattered and follow a straight line with slope -13.75$\pm$3.92 (Figure \ref{fig:median}). Due to the less scattering within the results from four bins, the estimated 95$\%$ confidence interval is very narrow compared with redshift results.
The confidence interval becomes narrower in the case of the Eddington ratio dependence of viewing angle. A straight line is fitted with slope -11.95$\pm$3.24 as shown in Figure \ref{fig:median}.
In all the plots along with 95$\%$ confidence level, we have also shown the 95$\%$ predicted interval, which says that the future observations will fall within this interval. As the plot shows, the viewing angle in each bin of redshift, mass, and Eddington ratio lies within the 95$\%$ confidence interval. Hence, we can say that there is no viewing angle evolution with respect to these source parameters. Our result suggests that there are no such biases in quasar; hence, they are very reliable candidates for cosmological purposes.
However, this result could be better constrained with more number of sources in our sample with simultaneous broadband observations. Since the quasars are variable, a simultaneity in observations is necessary in order to access the similar spectral property across the wavebands.

\subsection{Implications for the error margin in quasar-based cosmology}

\begin{figure}
    \centering
    \includegraphics[scale=0.45]{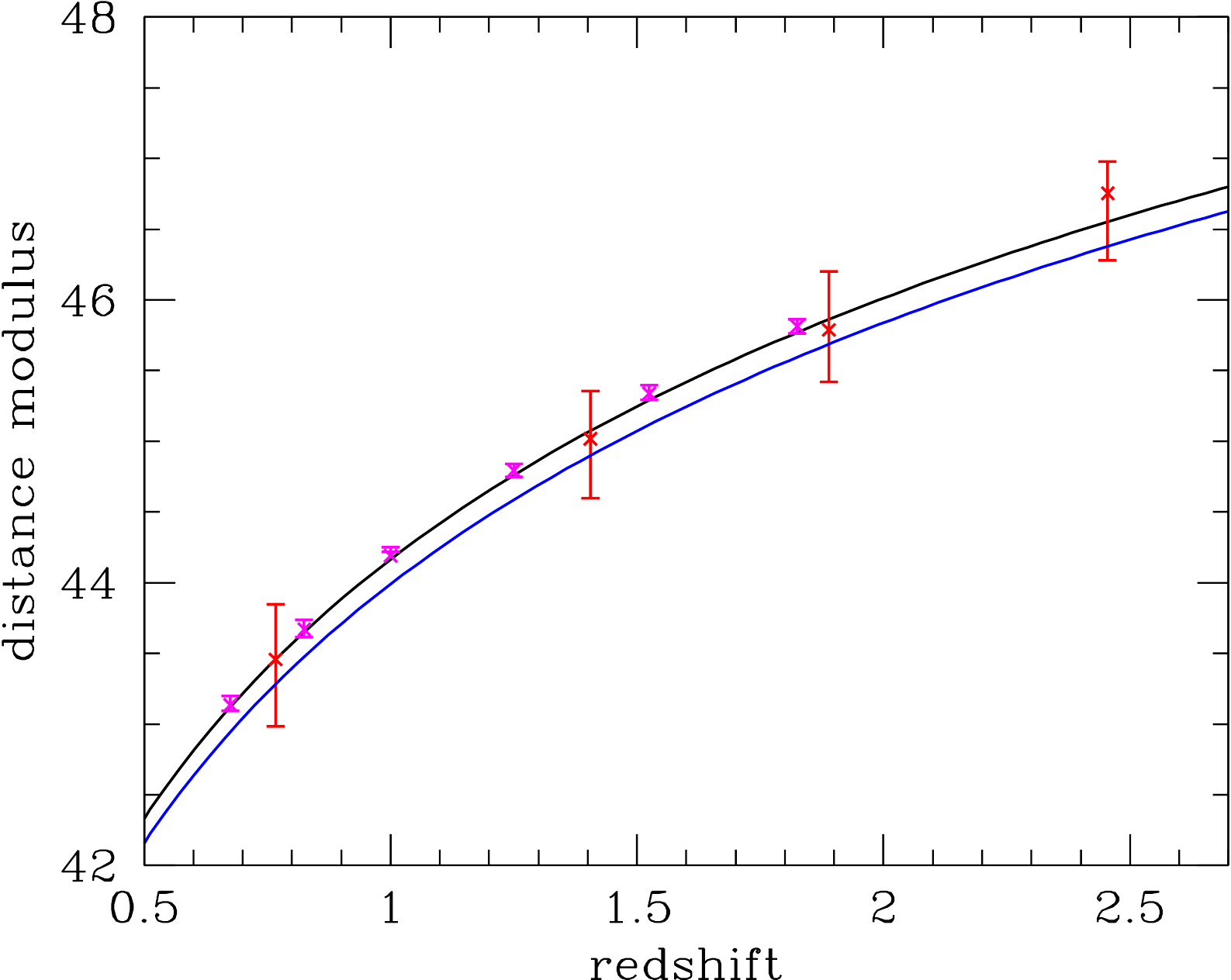}
    \caption{The distance modulus determination from the quasar sample when the viewing angle at the lowest redshift bin is artificially used for higher redshifts. Red points - our sample fitted by clumpy torus model, magenta points - sample of \citet{Gupta_2016}, and solid torus model. Black line is the standard $\Lambda$CDM cosmology from Planck \citep{Planck-Collaboration-2018-VI}, and the data points in the lowest redshift bin are calibrated there. None of the samples show a disagreement with the model at high redshifts. Blue line is the standard $\Lambda$CDM model, but with the Hubble constant $H_0$ from \citet{riess2019}. }
    \label{fig:cosmology}
\end{figure}

As we discussed in \citet{prince2021}, any systematic trends of the viewing angle with redshift may lead to incorrect cosmological predictions for the quasar-based methods which use the emission from flat accretion disks (continuum time-delay approach, \citealt{collier1999,cackett2007}, emission line time-delay approach, \citealt{watson2011,haas2011,czerny2013}, or UV/X-ray relation \citet{risaliti2015}. According to our results, the viewing angle in all redshift bins is consistent with the mean value of 47.4 deg, measured from the symmetry axis. The same value measured directly from the total distribution is equal to 53.6$^{+8.0}_{-16.0}$. 

However, to illustrate better the accuracy of our statement, we construct the Hubble diagram assuming that the viewing angle determination in the first redshift bin is correct. Then for the other redshift bins, the viewing angle determination was not performed. We do this exercise for a specific cosmological model taken from Planck results \citep{Planck-Collaboration-2018-VI}:$H_0 = 67.4$ km s$^{-1}$ Mpc$^{-1}$, $\Omega_m = 0.315$ (flat model).

The result is shown in Figure~\ref{fig:cosmology}. Red points reflect our viewing angles for four redshift bins, and the errors represent the consequence of the viewing angle uncertainty. The black line shows the standard Planck model, and it is comfortably consistent with the points within their errors. Therefore, quasar evolution with redshift does not seem to pose any problems for the tests of cosmology.

\section{Discussion}

In this paper, we did a careful search for the trends in the dusty torus properties with the redshift and the black hole mass and Eddington ratio.  Particularly the first trend is vital from the point of view of the quasar applications for cosmology. We use the advanced clumpy torus model of \citet{Stalevski_2012,Stalevski_2016} since only a clumpy material distribution is accounting properly for the broadband spectra \citep{nenkova2002,sebastian2006,markowitz2014,ogawa2021}. Our results show that no such trends are firmly detected in our sample of 80 objects, which implies that there is no need to introduce any systematic correction to the average orientations of quasars as a function of redshift since the obscuration by dusty torus does not change.

Our determination of the viewing angle has relatively large errors due to the model complexity. We also used only one model of the torus while more advanced models are currently available \cite[see e.g.][for their tests with the data]{gonzalez2019}, but the details are most important for high-quality data showing spectral features in the IR. With better quality data, better, fully self-consistent models should be used. The current model is based on optxagn of \citet{Chris_2012} which assumes the viewing angle of the corona/disk system at 60$^{\circ}$. This assumption is very good for the proper normalization of the incident continuum for the dusty torus radiative transfer since it represents well the bolometric luminosity of the source. Also the hot corona part and the warm corona emit roughly isotropically so the lack of explicit angle dependence in the optical band is expected only in the 6000 - 10000 \AA~ part of the spectrum, dominated by the standard disk. Since in our models the outer radius of the warm corona is located at $50 - 80 r_{g}$, most of the energy is dissipated in the hot and warm corona.  Therefore, this oversimplification should not affect our conclusions in the context of available data.

\citet{Alonso_Herrero_2011} have also used the \texttt{CLUMPY} torus model to fit the infrared spectral energy distribution in nearby Seyfert galaxies. They were able to constrain the torus parameters such as viewing angle, opening angle, number of clouds along the line of sight. They have also combined their Seyfert sample with the PG quasars from the literature to span the bolometric luminosity range from 10$^{43}$--10$^{47}$ erg/s and derived the torus geometrical covering factor, which highly depends on the bolometric luminosity. For the high luminous sources, they found the lower covering factor, and for the low luminosity sources (10$^{43}$--10$^{44}$), the covering factor was estimated as 0.9-1.0. The average viewing angle estimated in their paper is $\sim$ 49 degree (see Table 5 of \citealt{Alonso_Herrero_2011}), which is consistent within the errorbar with the estimated average viewing for our sample, i.e., 53.6$^{+8}_{-16}$. However, they could not study the redshift trend since their sample was limited to low redshift sources ($z < 0.016$). 

Alternatively, there are simpler ways to constrain the mean viewing angle of quasars as a function of redshift. They have based either on the measurements of Type 2 to Type 1 ratios in AGN samples or on measurements of the mid-IR to the bolometric luminosity ratios. Such measurements can be converted to the mean viewing angle under simplifying the assumption of the torus in the form of a solid body parameterized just be the opening angle. 
The ratios of the Type 2 to the total number of Type 1 + type 2 sources correspond to the solid angle (opening angle) of the torus, $\sigma$, measured from the equatorial plane. As shown by \citet{ELITZUR2008}, the fraction of Type 2 (f2) sources in the total population relates to $\sigma$ as f2 = sin $\sigma$, and the viewing angle for the type 1 sources would fall between 0 to $\uppi$/2 - $\sigma$, with the mean value of the viewing angle, $i_{eff}$ given by
\begin{equation}
\cos i_{eff} = 0.5(1 + \sin \sigma).
\end{equation}

 The second method was used by us in \citet{prince2021}, where we took the measurements of the IR to optical flux ratios from \citet{gu2013} which were in the original paper directly converted to covering factor:
\begin{equation}
    CF = L_{IR}/L_{bol}.
\end{equation}
\citet{gu2013} at the basis of this expression concluded about the time evolution of the torus covering factor, $CF$, with redshift. In \citet{prince2021} we used an improved 2-parameter model which for a single measured quantity (flux ratios) could not recover both parameters: torus opening angle and torus optical depth, which gave us broad limits in each of the two studied redshifts implying that the evolution may possibly be important.

These expressions have to be modified for the clumpy, soft-edge torus model. 
In the clumpy torus model, the classification of type 1 and type 2 AGN is no longer a matter of viewing angle but rather derived from the probability for the direct view of the AGN. Assuming clouds are optically thick, the probability for an AGN-produced photon escaping through the torus at viewing angle $i$ is P$_{esc}$ = $\simeq$e$^{-N0}$, where N0 is the number of clouds along the line of sight. In this case, the fraction of type 2 AGN is defined as f2 = (1 - e$^{-N0}$) sin $\sigma$, and it requires the additional knowledge to convert $f2$ to the mean viewing angle. 

Numerical torus models used in the current paper thus require measurements at various wavelengths in order to determine the model parameters. The advantage is that in principle we are able to determine all these parameters, but the data requirements are considerable, and the currently available data do not give results with high precision, as shown in Section~\ref{sect:results}. The analysis cannot be repeated for the same data as in \citet{gu2013} since for these objects we do not have broad band data coverage, including X-rays.

\citet{Treister2006} collected a sample of 2341 Hard X-ray sources and studied the evolution of the fraction of obscured AGN. Their sample covered the redshift range between 0-4. They found that there was no significant evolution of the fraction of obscured AGN with respect to redshift. However, they pointed out that considering the effect of spectroscopic incompleteness in the sample can change the results significantly. They have shown that in this case, the
intrinsic fraction of obscured AGN strongly depends on redshift and follows a relation (1+z)$^{\alpha}$, where $\alpha$ is 0.4$\pm$0.1. Later, a similar trend is also noticed in \citet{hasinger2008}, who performed large sample statistical studies based on a deep X-ray survey. They reported no redshift dependence of the Type 2 fraction up to the redshift $\sim 3$ when the direct sample was used. However, when supplementing with the possible bias, they have fitted the trend with two power-law, one below redshift 2.0 and the other above redshift 2.0. His best fit for the $\alpha$ below redshift 2.0 is 0.62$\pm$0.11 and above redshift 2.0 is 0.48$\pm$0.08.

A detailed study by \citet{Lawrence_2010} was done on samples selected from different wavebands (radio/IR/X-ray) to estimate the fraction f2 of type 2 AGN. Their result shows that the f2 is constant with bolometric luminosity for the sources selected based on radio and IR observations. However, similarly to \citet{hasinger2008}, they have also noticed strong dependence of f2 on bolometric luminosity for sources selected at the basis of the X-ray observations from Swift-BAT. To explain this strong dependence of f2 on X-ray luminosity, they have argued that many objects observed from the Swift-BAT are partially covered with Compton-thick material and partially covered by the material of intermediate thickness, and such objects would be seen apparently as Compton thin but with a suppressed X-ray luminosity, and hence this would cause an artificial correlation between obscured fraction and the X-ray luminosity.
A strong decreasing trend of dust covering factor with respect to bolometric luminosity was also seen in \citet{Maiolino2007}. Their sample consisted of 25 high luminosity quasars between redshift 2 -- 3.5 and local low luminosity type 1 AGN from the archival {\it Spitzer} IRS observations. The entire sample covers five orders of magnitude in the luminosity. These studies, however, did not report any specific trend with the redshift.

More studies used the ratio of the IR emission to the bolometric luminosity as a way to measure the torus covering factor (CF = L$_{IR}$/L$_{bol}$). \citet{lusso2013} reported weak dependence of the covering factor on the bolometric luminosity and no trends of the covering fraction with the redshift. However, \citet{gu2013} claimed a significant systematic difference in the covering factor between two selected redshift bins. 
He considered two population of sources: one in redshift bin $0.7\leq z \leq 1.1$ and other in redshift bin $2.0\leq z \leq2.4$.  
His results show the clear increasing trend with redshift, which suggests the change in dusty torus structure at higher redshift, and we used these data in our previous study of potential bias in quasar-based cosmology  \citep{prince2021}.

A detailed study on the evolution of covering factor in radio-quiet and the radio-loud source was recently done by \citet{Gupta_2016}. They had more number of sources than we have. Their sample had $\sim$800 radio-loud and a corresponding number of quiet radio quasars matched to the radio loud counterparts in the mass and the Eddington ratio, covering a wide range of redshift ($0\leq z \leq2$). The covering factor was estimated as the ratio of mid-infrared (MIR) luminosity to the bolometric luminosity (L$_{MIR}$/L$_{bol}$), assuming both radiations are isotropic and solid structure of dusty torus. They have studied the evolution of the covering factor of the whole sample with respect to mass, redshift, and Eddington ratio. They did not find any significant evolution with redshift claimed by \citet{gu2013}, and no evolution with the black hole mass nor with the Eddington ratio. Since they have a large sample, we used a solid torus model to convert their results to constraints for the evolution of the viewing angle with redshift. The solid torus model is certainly a gross oversimplification, but their results are quite consistent with our determination of the viewing angles from much better modeling. Thus we added these data to Figure~\ref{fig:cosmology}. The sample covers only redshifts up to 2, errors are very small, but again they do not predict any false redshift trends connected with the quasar evolution with redshift. 

In \citet{prince2021}, we discuss the possible error caused by the differences in the viewing angle when AGN are used for measurements of cosmological parameters at higher redshift since we do not know the exact structure of the torus. \citet{gu2013} suggested that the covering factor, which is the indirect representation of the torus structure changes with redshift which might imply that assuming the same effective viewing angle in time-delay measurement at all redshift might be not correct. The result was also consistent with our toy model that is presented in \citet{prince2021}. These two papers suggested a possibility that  the viewing angle evolves. Now, in the current paper we tried to directly measure the viewing angle of the individual quasars by SED modeling to check if such evolution indeed takes place. The result we got shows, however, that the viewing angle does not evolved with redshift, within the accuracy of our measurements and for the considered sample. Here could be two possibilities for these differences,
1) We need to have bigger, unbiased sample with a better broad-band spectral coverage to reduce error measurement and verify these conclusions
2) \citet{gu2013} results are biased by sample selection.

\section{Conclusions}

In this paper, we estimated the viewing angle of the individual objects from the broadband SED modeling covering the IR to X-ray range. After cross-matching the COSMOS and the SDSS DR14 catalog, we left with the total 90 sources which have broadband data, however, among those 10 sources have less data points than the total model parameters and hence not modeled in this study. To model the broadband SED, we have used the combination of \texttt{optxagnf} and the \texttt{SKIRTOR} models. Our main objective is to constrain the torus parameters (viewing angle, optical depth, and the opening angle). Individual modeling returns the viewing angle above 60$^\circ$ and the $\chi^2$-distributions shows a flat distribution and hence, we could not constrain the lower limit of viewing angle. To obtain the better constrain,
we have divided the 80 sources into four bins with respect to redshift, BH mass, and Eddington ratio. The probability distribution ($\sim$ exp($-\chi^2$)) of all sources in each bin are multiplied together to have cumulative probability distribution (shown in \textit{Appendix A}), which help us to constrain the lower and upper bound of the viewing angle. Further, we examined the evolution of viewing angle with these source parameters. The viewing angle of AGNs is the key parameter that can indirectly be used to constrain the cosmological parameters and recently discovered 'Hubble tension'.
We do not see a significant evolution of viewing angle with respect to redshift, BH mass, and Eddington ratio, and the Hubble diagram derived from this study is shown in the Figure \ref{fig:cosmology} with four red points corresponding to four redshift bins, which follows the standard $\Lambda$CDM model (black line). 
Therefore, We conclude that quasars are very good candidates for unbiased tracers of the Universe expansion.


\begin{center}
\begin{longtable*}[c]{cccccc|ccc}
\caption{The first six columns are the source parameters of our sample taken from the various catalog (see the Sample selection section) . The viewing angle, optical depth, and opening angle shown in last three columns are the estimated values from the broadband SED modeling. The optical depth is measured along the line of sight. }
\label{tab:my-table}\\
\hline\hline\noalign{\vskip 0.1cm}
SDSS Identifier & RA (deg) & Dec (deg) & z & log M$_{\rm{BH}}$ & log $\lambda_{\rm{Edd}}$& viewing & optical& opening\\
\endfirsthead
\endhead
 & (J2000) & (J2000) &  & (M$_{\odot}$)& & angle
 & depth &angle \\
 
 & & & & & &(degrees)&$(\tau)$&(degrees)\\
 
 \hline\hline\noalign{\vskip 0.1cm}
J095743.33+024823.7 & 149.4306 & 2.8066 & 1.362 & 8.426$\pm$0.288 & -0.899 & 60.8 & 0.002&10.0 \\
J095754.70+023832.7 & 149.4779 & 2.6424 & 1.600 & 8.853$\pm$0.099 & -0.703 & 69.6 & 0.140 & 10.0  \\
J095755.08+024806.3 & 149.4795 & 2.8018 & 1.111 & 8.697$\pm$0.153 & -0.825 & 55.2 & 0.052 & 15.04\\
J095756.64+020719.3 & 149.4860 & 2.1220 & 0.954 & 8.717$\pm$0.361 & -1.555 & 69.6 & 1.39 & 18.34\\
J095759.50+020435.9 & 149.4979 & 2.0766 & 2.034 & 9.495$\pm$0.064 & -0.754 & 69.6 & 0.047 & 10.0  \\
J095805.84+015158.9 & 149.5244 & 1.8664 & 0.660 & 7.978$\pm$0.201 & -1.477 &69.6&1.028&19.91\\
J095806.97+022248.4 & 149.5291 & 2.3801 & 3.108 & 6.862$\pm$10.796 & 1.059 &-&-&-\\
J095810.88+014005.1 & 149.5454 & 1.6681 & 2.101 & 8.615$\pm$0.048 & -0.364 & 64.0 & 0.013 & 10.0\\
J095815.50+014922.9 & 149.5646 & 1.8231 & 1.510 & 8.953$\pm$0.176 & -1.150 & 69.6 & 0.171 & 10.0\\
J095819.35+013530.5 & 149.5806 & 1.5918 & 3.055 & 7.502$\pm$0.059 & 0.319 &-&-&-\\
J095819.87+022903.5 & 149.5828 & 2.4843 & 0.345 & 8.453$\pm$0.112 & -1.487 & 74.4 & 0.965 & 10.0\\
J095820.44+020303.9 & 149.5852 & 2.0511 & 1.355 & 9.110$\pm$0.142 & -1.350 & 76.0 & 1.498 & 10.0\\
J095821.65+024628.1 & 149.5902 & 2.7745 & 1.405 & 9.467$\pm$0.071 & -1.463 & 71.2 & 0.321 & 10.0\\
J095822.18+014524.1 & 149.5924 & 1.7567 & 1.964 & 9.534$\pm$0.052 & -0.514 & 69.6 & 0.171 & 10.0\\
J095834.05+024427.1 & 149.6419 & 2.7409 & 1.887 & 9.071$\pm$0.113 & -0.842 & 69.6 & 0.6792& 12.69\\
J095834.74+014502.3 & 149.6448 & 1.7507 & 1.889 & 9.363$\pm$0.079 & -1.055 & 45.6 & 0.71E-7& 10.44\\
J095835.98+015157.0 & 149.6499 & 1.8658 & 2.932 & 8.271$\pm$0.469 & 0.291 & 52.8  & 0.726 & 22.56\\
J095844.94+014309.0 & 149.6873 & 1.7192 & 1.337 & 9.545$\pm$0.286 & -1.822 & 77.6 & 2.804 & 10.93\\
J095848.86+023441.1 & 149.7036 & 2.5781 & 1.549 & 8.629$\pm$0.135 & -1.048 & 69.6 & 0.078 & 10.0\\
J095852.15+025156.3 & 149.7173 & 2.8657 & 1.407 & 8.964$\pm$0.047 & -0.937 & 69.6 & 0.109 & 10.0\\
J095857.34+021314.5 & 149.7389 & 2.2207 & 1.024 & 9.332$\pm$0.067 & -1.959 & 78.4 & 0.979 & 10.0\\
J095858.67+020138.9 & 149.7445 & 2.0275 & 2.455 & 8.633$\pm$0.162 & 0.151  & 69.6 & 0.109 & 10.0\\
J095902.76+021906.3 & 149.7615 & 2.3184 & 0.345 & 8.791$\pm$0.253 & -1.903 & 73.6 & 0.747 & 10.0\\
J095907.65+020820.6 & 149.7819 & 2.1391 & 0.354 & 8.057$\pm$0.372 & -1.972 & 59.2 & 0.688E-4& 10.0\\
J095908.31+024309.6 & 149.7847 & 2.7193 & 1.318 & 9.346$\pm$0.170 & -1.090 & 69.6 & 0.078 & 10\\
J095908.68+025423.9 & 149.7862 & 2.9067 & 1.556 & 8.604$\pm$0.102 & -0.235 & 69.6 & 0.140 & 10.0\\
J095909.54+021916.4 & 149.7898 & 2.3213 & 0.378 & 7.301$\pm$0.235 & -1.006 & 75.2 & 1.231 & 10.0\\
J095918.70+020951.5 & 149.8279 & 2.1643 & 1.157 & 10.310$\pm$0.287 & -2.579 & 60.8 &0.594E-3& 10.0\\
J095919.82+024238.6 & 149.8326 & 2.7107 & 2.123 & 8.722$\pm$0.212 & -1.090 & 42.4 &2.127E-7& 11.36\\
J095923.55+022227.2 & 149.8481 & 2.3742 & 2.735 & 9.105$\pm$0.238 & -1.110 &-&-&-\\
J095924.46+015954.3 & 149.8519 & 1.9984 & 1.235 & 9.198$\pm$0.043 & -1.055 & 72.8 & 0.571& 10.0\\
J095930.20+024124.8 & 149.8759 & 2.6902 & 2.187 & 8.233$\pm$0.414 & -0.672 &-&-&-\\
J095935.67+024839.1 & 149.8986 & 2.8109 & 1.973 & 8.426$\pm$0.376 & -0.225 & 69.6 & 0.068 & 10.48\\
J095942.08+024103.1 & 149.9254 & 2.6842 & 1.780 & 9.081$\pm$0.259 & -0.983 & 69.6 & 0.047 & 10.0\\
J095943.41+020707.4 & 149.9309 & 2.1187 & 2.210 & 9.373$\pm$0.045 & -0.923 & 76.8 & 1.108 & 10.27\\
J095946.01+024743.5 & 149.9417 & 2.7955 & 1.067 & 8.730$\pm$0.049 & -0.735 & 69.6 & 0.1565& 10.0\\
J095949.39+020140.9 & 149.9558 & 2.0281 & 1.754 & 8.120$\pm$0.852 & 0.185  & 69.6 &0.047&10.0 \\
J095957.97+014327.4 & 149.9916 & 1.7243 & 1.627 & 8.919$\pm$0.283 & -1.066 & 35.2 &8.169E-13& 10.0\\
J100001.44+024844.6 & 150.0060 & 2.8124 & 0.766 & 8.482$\pm$0.231 & -0.915 & 69.6 & 0.171 & 10.0\\
J100008.14+013306.4 & 150.0339 & 1.5518 & 1.172 & 8.593$\pm$0.112 & -1.146 & 69.6 & 0.162 & 10.14\\
J100012.91+023522.8 & 150.0538 & 2.5897 & 0.698 & 8.137$\pm$0.182 & -0.415 & 69.6 & 0.047 & 10.0\\
J100013.54+013738.8 & 150.0564 & 1.6275 & 1.608 & 8.212$\pm$0.646 & -0.384 & 69.6 & 0.109 & 10.0\\
J100014.13+020054.4 & 150.0589 & 2.0152 & 2.498 & 8.629$\pm$0.043 & -0.062 & 61.6 & 0.942E-3&10.0 \\
J100024.38+015053.9 & 150.1016 & 1.8483 & 1.664 & 8.613$\pm$0.284 & -0.506 & 69.6 & 0.109 & 10.0\\
J100024.47+020619.6 & 150.1020 & 2.1055 & 2.288 & 8.934$\pm$0.052 & -0.417 & 66.4 & 0.984 & 15.19\\
J100024.64+023149.0 & 150.1027 & 2.5303 & 1.321 & 8.903$\pm$0.022 & -0.730 &-&-&-\\
J100024.88+023956.5 & 150.1037 & 2.6657 & 2.955 & 8.430$\pm$0.081 & -0.228 & 41.6 & 0.5652& 29.37\\
J100025.24+015852.0 & 150.1052 & 1.9811 & 0.373 & 8.264$\pm$0.115 & -1.132 & 69.6 & 0.047 & 10.0\\
J100028.63+025112.6 & 150.1193 & 2.8535 & 0.767 & 9.166$\pm$0.191 & -1.736 & 73.6 & 0.137 & 10.61\\
J100038.01+020822.4 & 150.1584 & 2.1396 & 1.828 & 9.049$\pm$0.216 & -1.120 & 64.8 & 0.107 & 11.70\\
J100043.13+020637.2 & 150.1797 & 2.1103 & 0.360 & 7.753$\pm$0.121 & -1.165 & 69.6 & 0.171 & 10.0\\
J100047.75+020756.9 & 150.1990 & 2.1325 & 2.153 & 8.603$\pm$0.114 & -0.559 & 63.2 & 0.003 & 10.0\\
J100049.91+020459.9 & 150.2080 & 2.0833 & 1.234 & 8.134$\pm$0.133 & -0.243 & 69.6 & 0.171 & 10.0\\
J100050.12+022854.8 & 150.2089 & 2.4819 & 3.365 & 9.467$\pm$0.137 & -0.931 & 56.8 & 1.991 & 26.36\\
J100050.15+022618.4 & 150.2090 & 2.4385 & 3.718 & 9.038$\pm$0.626 & -0.950 &-&-&-\\
J100055.39+023441.3 & 150.2308 & 2.5782 & 1.404 & 9.044$\pm$0.040 & -1.269 & 69.6 & 0.100 & 10.0\\
J100058.84+015400.2 & 150.2452 & 1.9000 & 1.557 & 8.743$\pm$0.156 & -0.934 & 72.8 & 0.156 & 10.0\\
J100105.31+021348.3 & 150.2721 & 2.2301 & 2.611 & 8.996$\pm$0.211 & -1.039 &-&-&-\\
J100110.19+023242.2 & 150.2925 & 2.5451 & 2.655 & 8.701$\pm$0.132 & -0.749 & 29.6 & 0.034 & 25.14\\
J100114.29+022356.8 & 150.3096 & 2.3991 & 1.796 & 9.187$\pm$0.131 & -0.909 & 57.6 & 0.304E-3& 10.0\\
J100115.96+021448.4 & 150.3165 & 2.2468 & 2.483 & 8.365$\pm$0.157 & -0.832 &-&-&-\\
J100116.78+014053.5 & 150.3200 & 1.6816 & 2.054 & 8.683$\pm$0.216 & -0.243 & 69.6 & 0.109 & 10.0\\
J100120.26+023341.3 & 150.3344 & 2.5615 & 1.834 & 8.625$\pm$0.508 & -0.722 & 69.6 & 0.063 & 10.38\\
J100123.01+020851.0 & 150.3459 & 2.1475 & 1.253 & 8.723$\pm$0.514 & -1.344 &-&-&-\\
J100130.36+014304.3 & 150.3766 & 1.7179 & 1.570 & 8.631$\pm$0.151 & -0.917 & 69.6 & 0.079 & 10.0\\
J100132.16+013420.8 & 150.3840 & 1.5724 & 1.355 & 8.255$\pm$0.133 & -0.557 & 69.6 & 0.171 & 10.0\\
J100136.50+025303.6 & 150.4021 & 2.8843 & 2.116 & 8.810$\pm$0.047 & -0.304 & 61.6 & 0.003 & 10.0\\
J100140.95+015643.1 & 150.4207 & 1.9453 & 2.180 & 8.318$\pm$0.093 & 0.256  &-&-&-\\
J100145.15+022456.9 & 150.4381 & 2.4158 & 2.032 & 8.851$\pm$0.534 & -0.742 & 75.2 & 1.006 & 10.0\\
J100157.77+024631.6 & 150.4907 & 2.7755 & 1.434 & 9.283$\pm$0.050 & -1.611 & 77.6 & 0.645 & 10.0\\
J100159.78+022641.7 & 150.4991 & 2.4449 & 2.023 & 8.688$\pm$0.035 & -0.274 &59.55&0.019&12.88\\
J100201.50+020329.4 & 150.5063 & 2.0582 & 2.023 & 9.074$\pm$0.295 & -0.455 & 69.6 & 0.078 & 10.0\\
J100202.77+022434.6 & 150.5116 & 2.4096 & 0.988 & 9.085$\pm$0.136 & -1.637 & 72.0 & 0.945 & 11.99\\
J100210.72+023026.2 & 150.5447 & 2.5073 & 1.161 & 8.264$\pm$0.149 & -0.515 & 69.6 & 0.083 & 10.0\\
J100219.49+015537.0 & 150.5812 & 1.9269 & 1.510 & 9.070$\pm$0.126 & -1.299 & 69.6 & 0.171 & 10.0\\
J100226.33+021923.1 & 150.6097 & 2.3231 & 1.292 & 8.311$\pm$0.154 & -0.657 & 69.6 & 0.171 & 10.0\\
J100232.13+023537.3 & 150.6339 & 2.5937 & 0.658 & 8.246$\pm$0.868 & -0.630 & 69.6 & 0.171 & 10.0\\
J100234.32+015011.1 & 150.6431 & 1.8365 & 1.505 & 8.747$\pm$0.044 & -0.616 & 69.6 & 0.171 & 10.0\\
J100236.70+015948.6 & 150.6529 & 1.9968 & 1.516 & 8.263$\pm$0.396 & -0.534 & 69.6 & 0.171 & 10.0\\
J100237.91+024700.7 & 150.6580 & 2.7836 & 0.212 & 8.866$\pm$0.228 & -2.573 & 60.0 & 4.430 & 31.45\\
J100238.25+013746.5 & 150.6594 & 1.6296 & 2.506 & 9.359$\pm$0.024 & -0.486 & 71.2 & 0.206 & 10.0\\
J100238.70+013936.4 & 150.6613 & 1.6601 & 1.318 & 8.691$\pm$0.097 & -1.394 & 53.6 & 0.003 & 12.65\\
J100249.32+023746.5 & 150.7055 & 2.6296 & 2.130 & 9.325$\pm$0.373 & -1.179 & 69.6 & 0.171 & 10.0\\
J100251.62+022905.3 & 150.7151 & 2.4848 & 2.006 & 9.022$\pm$0.023 & -0.536 &54.4&0.004&12.676\\
J100256.53+021158.4 & 150.7356 & 2.1996 & 3.503 & 9.080$\pm$0.197 & -0.712 & 48.0 & 0.428 & 25.12\\
J100302.91+015208.5 & 150.7622 & 1.8690 & 1.799 & 8.794$\pm$0.094 & -0.639 & 74.4 & 0.263 & 10.0\\
J100309.21+022038.3 & 150.7884 & 2.3440 & 1.965 & 8.530$\pm$0.107 & -0.380 & 32.0 & 2.498E-14&10.0\\
J100320.90+022930.0 & 150.8371 & 2.4917 & 4.412 & 8.547$\pm$0.337 & -0.245 & 32.8 & 0.907 & 39.99\\
J100324.56+021831.3 & 150.8524 & 2.3087 & 0.518 & 8.623$\pm$0.081 & -1.007 & 69.6 & 0.109 & 10.0\\
J100327.67+015742.3 & 150.8654 & 1.9618 & 1.067 & 8.640$\pm$0.066 & -0.475 & 69.6 & 0.079 & 10.0\\
\hline
\end{longtable*}
\end{center}

\footnotesize{{\sc NOTES.} Columns are as follows: (1) Object name, (2) right ascension, (3) declination, (4) redshift, (5) fiducial BH mass (in log-scale), (6) Eddington ratio (in log-scale), (7) viewing angle to the distant observer (wrt the black hole spin axis), (8) the optical depth along the line of sight, and (9) the torus opening angle. Columns (5) and (6) are taken from the DR14 Quasar Catalogue \citep[][]{rakshitetal2019}. Columns (7)-(9) are estimated from our model described in Section \ref{sec3}.}

\begin{table*}[hbt!]
    \centering
    \begin{tabular}{ccccccc}
    \hline
redshift-bin& median-redshift&viewing angle (deg) & -ve error(deg) & +ve error(deg)\\
     1& 0.767    & 48.8 & 29.6 & 16.0 \\
     2& 1.405    & 51.2 & 20.0 & 13.6 \\
     3& 1.889    & 52.0 & 26.4 & 12.0 \\
     4 & 2.455   & 37.6 & 24.8 & 21.6 \\
     \hline
mass-bin& median-log(BH mass)&viewing angle (deg) & -ve error(deg) & +ve error(deg)\\
     1& 8.264    & 51.2 & 29.6 & 15.2 \\
     2& 8.629    & 41.6 & 22.4 & 16.8 \\
     3& 8.865    & 43.2 & 28.0 & 17.6 \\
     4& 9.339    & 35.2 & 24.8 & 23.2 \\    
     \hline
edd ratio-bin& median-eddington ratio&viewing angle (deg) & -ve error(deg) & +ve error(deg)\\
     1& 0.0373    & 44.8 & 23.2 & 18.4 \\
     2& 0.1097   & 41.6 & 25.6 & 17.6 \\
     3& 0.2297    & 41.6 & 28.0 & 17.6 \\
     4& 0.5703    & 37.6 & 26.4 & 23.2 \\    
     \hline
    \end{tabular}
    \caption{The average viewing angle estimated in each redshift, mass, and eddington ratio bins. }
    \label{tab:result}
\end{table*}

\acknowledgments
The project was partially supported by the Polish Funding Agency National Science Centre, project 2017/26/\-A/ST9/\-00756 (MAESTRO  9) and MNiSW grant DIR/WK/2018/12.
\\
\software {\textmyfont{MATPLOTLIB}  (\citealt{hunter07}); \textmyfont{NUMPY} (\citealt{numpy}); \textmyfont{TOPCAT} (\citealt{Taylor})}

\bibliography{main}

\begin{thebibliography}{}
\expandafter\ifx\csname natexlab\endcsname\relax\def\natexlab#1{#1}\fi
\providecommand{\url}[1]{\href{#1}{#1}}
\providecommand{\dodoi}[1]{doi:~\href{http://doi.org/#1}{\nolinkurl{#1}}}
\providecommand{\doeprint}[1]{\href{http://ascl.net/#1}{\nolinkurl{http://ascl.net/#1}}}
\providecommand{\doarXiv}[1]{\href{https://arxiv.org/abs/#1}{\nolinkurl{https://arxiv.org/abs/#1}}}

\bibitem[{Alonso-Herrero {et~al.}(2011)Alonso-Herrero, Almeida, Mason, Ramos,
  Roche, Levenson, Elitzur, Packham, Espinosa, Young, D{\'{\i}}az-Santos, \&
  P{\'{e}}rez-Garc{\'{\i}}a}]{Alonso_Herrero_2011}
Alonso-Herrero, A., Almeida, C.~R., Mason, R., {et~al.} 2011, The Astrophysical
  Journal, 736, 82, \dodoi{10.1088/0004-637x/736/2/82}

\bibitem[{{Arnaud}(1996)}]{xspec_1996}
{Arnaud}, K.~A. 1996, in Astronomical Society of the Pacific Conference Series,
  Vol. 101, Astronomical Data Analysis Software and Systems V, ed. G.~H.
  {Jacoby} \& J.~{Barnes}, 17

\bibitem[{{Assef} {et~al.}(2013){Assef}, {Stern}, {Kochanek}, {Blain},
  {Brodwin}, {Brown}, {Donoso}, {Eisenhardt}, {Jannuzi}, {Jarrett}, {Stanford},
  {Tsai}, {Wu}, \& {Yan}}]{Assef_2013}
{Assef}, R.~J., {Stern}, D., {Kochanek}, C.~S., {et~al.} 2013, \apj, 772, 26,
  \dodoi{10.1088/0004-637X/772/1/26}

\bibitem[{Bassett \& Hlozek(2010)}]{bassett_hlozek_2010}
Bassett, B., \& Hlozek, R. 2010, Baryon acoustic oscillations, ed.
  P.~Ruiz-Lapuente (Cambridge University Press), 246–278,
  \dodoi{10.1017/CBO9781139193627.010}

\bibitem[{{Becker} {et~al.}(1995){Becker}, {White}, \& {Helfand}}]{becker+95}
{Becker}, R.~H., {White}, R.~L., \& {Helfand}, D.~J. 1995, \apj, 450, 559,
  \dodoi{10.1086/176166}

\bibitem[{{Cackett} {et~al.}(2007){Cackett}, {Horne}, \&
  {Winkler}}]{cackett2007}
{Cackett}, E.~M., {Horne}, K., \& {Winkler}, H. 2007, \mnras, 380, 669,
  \dodoi{10.1111/j.1365-2966.2007.12098.x}

\bibitem[{{Civano} {et~al.}(2016){Civano}, {Marchesi}, {Comastri}, {Urry},
  {Elvis}, {Cappelluti}, {Puccetti}, {Brusa}, {Zamorani}, {Hasinger},
  {Aldcroft}, {Alexand er}, {Allevato}, {Brunner}, {Capak}, {Finoguenov},
  {Fiore}, {Fruscione}, {Gilli}, {Glotfelty}, {Griffiths}, {Hao}, {Harrison},
  {Jahnke}, {Kartaltepe}, {Karim}, {LaMassa}, {Lanzuisi}, {Miyaji}, {Ranalli},
  {Salvato}, {Sargent}, {Scoville}, {Schawinski}, {Schinnerer}, {Silverman},
  {Smolcic}, {Stern}, {Toft}, {Trakhtenbrot}, {Treister}, \&
  {Vignali}}]{civano+16}
{Civano}, F., {Marchesi}, S., {Comastri}, A., {et~al.} 2016, \apj, 819, 62,
  \dodoi{10.3847/0004-637X/819/1/62}

\bibitem[{{Collier} {et~al.}(1999){Collier}, {Horne}, {Wanders}, \&
  {Peterson}}]{collier1999}
{Collier}, S., {Horne}, K., {Wanders}, I., \& {Peterson}, B.~M. 1999, \mnras,
  302, L24, \dodoi{10.1046/j.1365-8711.1999.02250.x}

\bibitem[{{Cutri} {et~al.}(2003){Cutri}, {Skrutskie}, {van Dyk}, {Beichman},
  {Carpenter}, {Chester}, {Cambresy}, {Evans}, {Fowler}, {Gizis}, {Howard},
  {Huchra}, {Jarrett}, {Kopan}, {Kirkpatrick}, {Light}, {Marsh}, {McCallon},
  {Schneider}, {Stiening}, {Sykes}, {Weinberg}, {Wheaton}, {Wheelock}, \&
  {Zacarias}}]{cutri+03}
{Cutri}, R.~M., {Skrutskie}, M.~F., {van Dyk}, S., {et~al.} 2003, VizieR Online
  Data Catalog, II/246

\bibitem[{{Czerny} {et~al.}(2013){Czerny}, {Hryniewicz}, {Maity},
  {Schwarzenberg-Czerny}, {{\.Z}ycki}, \& {Bilicki}}]{czerny2013}
{Czerny}, B., {Hryniewicz}, K., {Maity}, I., {et~al.} 2013, \aap, 556, A97,
  \dodoi{10.1051/0004-6361/201220832}

\bibitem[{{Done} {et~al.}(2012){Done}, {Davis}, {Jin}, {Blaes}, \&
  {Ward}}]{Chris_2012}
{Done}, C., {Davis}, S.~W., {Jin}, C., {Blaes}, O., \& {Ward}, M. 2012, \mnras,
  420, 1848, \dodoi{10.1111/j.1365-2966.2011.19779.x}

\bibitem[{Elitzur(2008)}]{ELITZUR2008}
Elitzur, M. 2008, New Astronomy Reviews, 52, 274,
  \dodoi{https://doi.org/10.1016/j.newar.2008.06.010}

\bibitem[{{Elvis} {et~al.}(2009){Elvis}, {Civano}, {Vignali}, {Puccetti},
  {Fiore}, {Cappelluti}, {Aldcroft}, {Fruscione}, {Zamorani}, {Comastri},
  {Brusa}, {Gilli}, {Miyaji}, {Damiani}, {Koekemoer}, {Finoguenov}, {Brunner},
  {Urry}, {Silverman}, {Mainieri}, {Hasinger}, {Griffiths}, {Carollo}, {Hao},
  {Guzzo}, {Blain}, {Calzetti}, {Carilli}, {Capak}, {Ettori}, {Fabbiano},
  {Impey}, {Lilly}, {Mobasher}, {Rich}, {Salvato}, {Sand ers}, {Schinnerer},
  {Scoville}, {Shopbell}, {Taylor}, {Taniguchi}, \& {Volonteri}}]{elvis+09}
{Elvis}, M., {Civano}, F., {Vignali}, C., {et~al.} 2009, \apjs, 184, 158,
  \dodoi{10.1088/0067-0049/184/1/158}

\bibitem[{{Freedman} {et~al.}(2019){Freedman}, {Madore}, {Hatt}, {Hoyt},
  {Jang}, {Beaton}, {Burns}, {Lee}, {Monson}, {Neeley}, {Phillips}, {Rich}, \&
  {Seibert}}]{freedman2019}
{Freedman}, W.~L., {Madore}, B.~F., {Hatt}, D., {et~al.} 2019, \apj, 882, 34,
  \dodoi{10.3847/1538-4357/ab2f73}

\bibitem[{{Gonz{\'a}lez-Mart{\'\i}n} {et~al.}(2019){Gonz{\'a}lez-Mart{\'\i}n},
  {Masegosa}, {Garc{\'\i}a-Bernete}, {Ramos Almeida},
  {Rodr{\'\i}guez-Espinosa}, {M{\'a}rquez}, {Esparza-Arredondo},
  {Osorio-Clavijo}, {Mart{\'\i}nez-Paredes}, {Victoria-Ceballos}, {Pasetto}, \&
  {Dultzin}}]{gonzalez2019}
{Gonz{\'a}lez-Mart{\'\i}n}, O., {Masegosa}, J., {Garc{\'\i}a-Bernete}, I.,
  {et~al.} 2019, \apj, 884, 11, \dodoi{10.3847/1538-4357/ab3e4f}

\bibitem[{{Gu}(2013)}]{gu2013}
{Gu}, M. 2013, \apj, 773, 176, \dodoi{10.1088/0004-637X/773/2/176}

\bibitem[{{Gupta} {et~al.}(2016){Gupta}, {Sikora}, \& {Nalewajko}}]{Gupta_2016}
{Gupta}, M., {Sikora}, M., \& {Nalewajko}, K. 2016, \mnras, 461, 2346,
  \dodoi{10.1093/mnras/stw1473}

\bibitem[{{Haas} {et~al.}(2011){Haas}, {Chini}, {Ramolla}, {Pozo Nu{\~n}ez},
  {Westhues}, {Watermann}, {Hoffmeister}, \& {Murphy}}]{haas2011}
{Haas}, M., {Chini}, R., {Ramolla}, M., {et~al.} 2011, \aap, 535, A73,
  \dodoi{10.1051/0004-6361/201117325}

\bibitem[{{Hasinger}(2008)}]{hasinger2008}
{Hasinger}, G. 2008, \aap, 490, 905, \dodoi{10.1051/0004-6361:200809839}

\bibitem[{{H{\"o}nig} {et~al.}(2006){H{\"o}nig}, {Beckert}, {Ohnaka}, \&
  {Weigelt}}]{sebastian2006}
{H{\"o}nig}, S.~F., {Beckert}, T., {Ohnaka}, K., \& {Weigelt}, G. 2006, \aap,
  452, 459, \dodoi{10.1051/0004-6361:20054622}

\bibitem[{{Hunter}(2007)}]{hunter07}
{Hunter}, J.~D. 2007, Computing in Science and Engineering, 9, 90,
  \dodoi{10.1109/MCSE.2007.55}

\bibitem[{{Lawrence} \& {Elvis}(2010)}]{Lawrence_2010}
{Lawrence}, A., \& {Elvis}, M. 2010, \apj, 714, 561,
  \dodoi{10.1088/0004-637X/714/1/561}

\bibitem[{{Lawrence} {et~al.}(2007){Lawrence}, {Warren}, {Almaini}, {Edge},
  {Hambly}, {Jameson}, {Lucas}, {Casali}, {Adamson}, {Dye}, {Emerson},
  {Foucaud}, {Hewett}, {Hirst}, {Hodgkin}, {Irwin}, {Lodieu}, {McMahon},
  {Simpson}, {Smail}, {Mortlock}, \& {Folger}}]{lawrence+07}
{Lawrence}, A., {Warren}, S.~J., {Almaini}, O., {et~al.} 2007, \mnras, 379,
  1599, \dodoi{10.1111/j.1365-2966.2007.12040.x}

\bibitem[{{Lusso} {et~al.}(2013){Lusso}, {Hennawi}, {Comastri}, {Zamorani},
  {Richards}, {Vignali}, {Treister}, {Schawinski}, {Salvato}, \&
  {Gilli}}]{lusso2013}
{Lusso}, E., {Hennawi}, J.~F., {Comastri}, A., {et~al.} 2013, \apj, 777, 86,
  \dodoi{10.1088/0004-637X/777/2/86}

\bibitem[{{Maiolino} {et~al.}(2007){Maiolino}, {Shemmer}, {Imanishi}, {Netzer},
  {Oliva}, {Lutz}, \& {Sturm}}]{Maiolino2007}
{Maiolino}, R., {Shemmer}, O., {Imanishi}, M., {et~al.} 2007, \aap, 468, 979,
  \dodoi{10.1051/0004-6361:20077252}

\bibitem[{{Marchesi} {et~al.}(2016){Marchesi}, {Civano}, {Elvis}, {Salvato},
  {Brusa}, {Comastri}, {Gilli}, {Hasinger}, {Lanzuisi}, {Miyaji}, {Treister},
  {Urry}, {Vignali}, {Zamorani}, {Allevato}, {Cappelluti}, {Cardamone},
  {Finoguenov}, {Griffiths}, {Karim}, {Laigle}, {LaMassa}, {Jahnke}, {Ranalli},
  {Schawinski}, {Schinnerer}, {Silverman}, {Smolcic}, {Suh}, \&
  {Trakhtenbrot}}]{marchesi+16}
{Marchesi}, S., {Civano}, F., {Elvis}, M., {et~al.} 2016, \apj, 817, 34,
  \dodoi{10.3847/0004-637X/817/1/34}

\bibitem[{{Markowitz} {et~al.}(2014){Markowitz}, {Krumpe}, \&
  {Nikutta}}]{markowitz2014}
{Markowitz}, A.~G., {Krumpe}, M., \& {Nikutta}, R. 2014, \mnras, 439, 1403,
  \dodoi{10.1093/mnras/stt2492}

\bibitem[{{Martin} {et~al.}(2005){Martin}, {Fanson}, {Schiminovich},
  {Morrissey}, {Friedman}, {Barlow}, {Conrow}, {Grange}, {Jelinsky},
  {Milliard}, {Siegmund}, {Bianchi}, {Byun}, {Donas}, {Forster}, {Heckman},
  {Lee}, {Madore}, {Malina}, {Neff}, {Rich}, {Small}, {Surber}, {Szalay},
  {Welsh}, \& {Wyder}}]{martin+05}
{Martin}, D.~C., {Fanson}, J., {Schiminovich}, D., {et~al.} 2005, \apjl, 619,
  L1, \dodoi{10.1086/426387}

\bibitem[{{Nenkova} {et~al.}(2002){Nenkova}, {Ivezi{\'c}}, \&
  {Elitzur}}]{nenkova2002}
{Nenkova}, M., {Ivezi{\'c}}, {\v{Z}}., \& {Elitzur}, M. 2002, \apjl, 570, L9,
  \dodoi{10.1086/340857}

\bibitem[{{Ogawa} {et~al.}(2021){Ogawa}, {Ueda}, {Tanimoto}, \&
  {Yamada}}]{ogawa2021}
{Ogawa}, S., {Ueda}, Y., {Tanimoto}, A., \& {Yamada}, S. 2021, \apj, 906, 84,
  \dodoi{10.3847/1538-4357/abccce}

\bibitem[{Oliphant(2015)}]{numpy}
Oliphant, T. 2015, {NumPy}: A guide to {NumPy}, 2nd edn., USA: CreateSpace
  Independent Publishing Platform.
\newblock \url{http://www.numpy.org/}

\bibitem[{{Palmese} {et~al.}(2019){Palmese}, {Graur}, {Annis}, {BenZvi}, {Di
  Valentino}, {Garcia-Bellido}, {Gontcho}, {Keeley}, {Kim}, {Lahav},
  {Nissanke}, {Paterson}, {Sako}, {Shafieloo}, \& {Tsai}}]{2019BAAS...51c.310P}
{Palmese}, A., {Graur}, O., {Annis}, J.~T., {et~al.} 2019, \baas, 51, 310.
\newblock \doarXiv{1903.04730}

\bibitem[{{P{\^a}ris} {et~al.}(2018){P{\^a}ris}, {Petitjean}, {Aubourg},
  {Myers}, {Streblyanska}, {Lyke}, {Anderson}, {Armengaud}, {Bautista},
  {Blanton}, {Blomqvist}, {Brinkmann}, {Brownstein}, {Brand t}, {Burtin},
  {Dawson}, {de la Torre}, {Georgakakis}, {Gil-Mar{\'\i}n}, {Green}, {Hall},
  {Kneib}, {LaMassa}, {Le Goff}, {MacLeod}, {Mariappan}, {McGreer}, {Merloni},
  {Noterdaeme}, {Palanque-Delabrouille}, {Percival}, {Ross}, {Rossi},
  {Schneider}, {Seo}, {Tojeiro}, {Weaver}, {Weijmans}, {Y{\`e}che}, {Zarrouk},
  \& {Zhao}}]{paris+18}
{P{\^a}ris}, I., {Petitjean}, P., {Aubourg}, {\'E}., {et~al.} 2018, \aap, 613,
  A51, \dodoi{10.1051/0004-6361/201732445}

\bibitem[{{Pesce} {et~al.}(2020){Pesce}, {Braatz}, {Reid}, {Riess}, {Scolnic},
  {Condon}, {Gao}, {Henkel}, {Impellizzeri}, {Kuo}, \&
  {Lo}}]{Pesce_2020_MegaMaser}
{Pesce}, D.~W., {Braatz}, J.~A., {Reid}, M.~J., {et~al.} 2020, \apjl, 891, L1,
  \dodoi{10.3847/2041-8213/ab75f0}

\bibitem[{{Planck Collaboration} {et~al.}(2020){Planck Collaboration},
  {Aghanim}, {Akrami}, {Ashdown}, {Aumont}, {Baccigalupi}, {Ballardini},
  {Banday}, {Barreiro}, {Bartolo}, {Basak}, {Battye}, {Benabed}, {Bernard},
  {Bersanelli}, {Bielewicz}, {Bock}, {Bond}, {Borrill}, {Bouchet}, {Boulanger},
  {Bucher}, {Burigana}, {Butler}, {Calabrese}, {Cardoso}, {Carron},
  {Challinor}, {Chiang}, {Chluba}, {Colombo}, {Combet}, {Contreras}, {Crill},
  {Cuttaia}, {de Bernardis}, {de Zotti}, {Delabrouille}, {Delouis}, {Di
  Valentino}, {Diego}, {Dor{\'e}}, {Douspis}, {Ducout}, {Dupac}, {Dusini},
  {Efstathiou}, {Elsner}, {En{\ss}lin}, {Eriksen}, {Fantaye}, {Farhang},
  {Fergusson}, {Fernandez-Cobos}, {Finelli}, {Forastieri}, {Frailis},
  {Fraisse}, {Franceschi}, {Frolov}, {Galeotta}, {Galli}, {Ganga},
  {G{\'e}nova-Santos}, {Gerbino}, {Ghosh}, {Gonz{\'a}lez-Nuevo}, {G{\'o}rski},
  {Gratton}, {Gruppuso}, {Gudmundsson}, {Hamann}, {Handley}, {Hansen},
  {Herranz}, {Hildebrandt}, {Hivon}, {Huang}, {Jaffe}, {Jones}, {Karakci},
  {Keih{\"a}nen}, {Keskitalo}, {Kiiveri}, {Kim}, {Kisner}, {Knox},
  {Krachmalnicoff}, {Kunz}, {Kurki-Suonio}, {Lagache}, {Lamarre}, {Lasenby},
  {Lattanzi}, {Lawrence}, {Le Jeune}, {Lemos}, {Lesgourgues}, {Levrier},
  {Lewis}, {Liguori}, {Lilje}, {Lilley}, {Lindholm}, {L{\'o}pez-Caniego},
  {Lubin}, {Ma}, {Mac{\'\i}as-P{\'e}rez}, {Maggio}, {Maino}, {Mandolesi},
  {Mangilli}, {Marcos-Caballero}, {Maris}, {Martin}, {Martinelli},
  {Mart{\'\i}nez-Gonz{\'a}lez}, {Matarrese}, {Mauri}, {McEwen}, {Meinhold},
  {Melchiorri}, {Mennella}, {Migliaccio}, {Millea}, {Mitra},
  {Miville-Desch{\^e}nes}, {Molinari}, {Montier}, {Morgante}, {Moss}, {Natoli},
  {N{\o}rgaard-Nielsen}, {Pagano}, {Paoletti}, {Partridge}, {Patanchon},
  {Peiris}, {Perrotta}, {Pettorino}, {Piacentini}, {Polastri}, {Polenta},
  {Puget}, {Rachen}, {Reinecke}, {Remazeilles}, {Renzi}, {Rocha}, {Rosset},
  {Roudier}, {Rubi{\~n}o-Mart{\'\i}n}, {Ruiz-Granados}, {Salvati}, {Sandri},
  {Savelainen}, {Scott}, {Shellard}, {Sirignano}, {Sirri}, {Spencer},
  {Sunyaev}, {Suur-Uski}, {Tauber}, {Tavagnacco}, {Tenti}, {Toffolatti},
  {Tomasi}, {Trombetti}, {Valenziano}, {Valiviita}, {Van Tent}, {Vibert},
  {Vielva}, {Villa}, {Vittorio}, {Wandelt}, {Wehus}, {White}, {White},
  {Zacchei}, \& {Zonca}}]{Planck-Collaboration-2018-VI}
{Planck Collaboration}, {Aghanim}, N., {Akrami}, Y., {et~al.} 2020, \aap, 641,
  A6, \dodoi{10.1051/0004-6361/201833910}

\bibitem[{{Prince} {et~al.}(2021){Prince}, {Czerny}, \& {Pollo}}]{prince2021}
{Prince}, R., {Czerny}, B., \& {Pollo}, A. 2021, arXiv e-prints,
  arXiv:2101.01244.
\newblock \doarXiv{2101.01244}

\bibitem[{{Rakshit} {et~al.}(2020){Rakshit}, {Stalin}, \&
  {Kotilainen}}]{rakshitetal2019}
{Rakshit}, S., {Stalin}, C.~S., \& {Kotilainen}, J. 2020, \apjs, 249, 17,
  \dodoi{10.3847/1538-4365/ab99c5}

\bibitem[{{Riess}(2019)}]{Riess2019NatRP}
{Riess}, A.~G. 2019, Nature Reviews Physics, 2, 10,
  \dodoi{10.1038/s42254-019-0137-0}

\bibitem[{{Riess} {et~al.}(2020){Riess}, {Casertano}, {Yuan}, {Bowers},
  {Macri}, {Zinn}, \& {Scolnic}}]{riess2020}
{Riess}, A.~G., {Casertano}, S., {Yuan}, W., {et~al.} 2020, arXiv e-prints,
  arXiv:2012.08534.
\newblock \doarXiv{2012.08534}

\bibitem[{{Riess} {et~al.}(2019){Riess}, {Casertano}, {Yuan}, {Macri}, \&
  {Scolnic}}]{riess2019}
{Riess}, A.~G., {Casertano}, S., {Yuan}, W., {Macri}, L.~M., \& {Scolnic}, D.
  2019, \apj, 876, 85, \dodoi{10.3847/1538-4357/ab1422}

\bibitem[{{Risaliti} \& {Lusso}(2015)}]{risaliti2015}
{Risaliti}, G., \& {Lusso}, E. 2015, \apj, 815, 33,
  \dodoi{10.1088/0004-637X/815/1/33}

\bibitem[{{Risaliti} \& {Lusso}(2019)}]{2019NatAs...3..272R}
---. 2019, Nature Astronomy, 3, 272, \dodoi{10.1038/s41550-018-0657-z}

\bibitem[{{Rosen} {et~al.}(2016){Rosen}, {Webb}, {Watson}, {Ballet}, {Barret},
  {Braito}, {Carrera}, {Ceballos}, {Coriat}, {Della Ceca}, {Denkinson},
  {Esquej}, {Farrell}, {Freyberg}, {Grise}, {Guillout}, {Heil}, {Law-Green},
  {Lamer}, {Lin}, {Martino}, {Michel}, {Motch}, {Nebot Gomez-Moran}, {Page},
  {Page}, {Page}, {Pakull}, {Pye}, {Read}, {Rodriguez}, {Sakano}, {Saxton},
  {Schwope}, {Scott}, {Sturm}, {Traulsen}, {Yershov}, \&
  {Zolotukhin}}]{rosen+16}
{Rosen}, S.~R., {Webb}, N.~A., {Watson}, M.~G., {et~al.} 2016, VizieR Online
  Data Catalog, IX/50

\bibitem[{Schaefer(2003)}]{Schaefer_2003}
Schaefer, B.~E. 2003, The Astrophysical Journal, 583, L67,
  \dodoi{10.1086/368104}

\bibitem[{{Skrutskie} {et~al.}(2006){Skrutskie}, {Cutri}, {Stiening},
  {Weinberg}, {Schneider}, {Carpenter}, {Beichman}, {Capps}, {Chester},
  {Elias}, {Huchra}, {Liebert}, {Lonsdale}, {Monet}, {Price}, {Seitzer},
  {Jarrett}, {Kirkpatrick}, {Gizis}, {Howard}, {Evans}, {Fowler}, {Fullmer},
  {Hurt}, {Light}, {Kopan}, {Marsh}, {McCallon}, {Tam}, {Van Dyk}, \&
  {Wheelock}}]{skrutskie+06}
{Skrutskie}, M.~F., {Cutri}, R.~M., {Stiening}, R., {et~al.} 2006, \aj, 131,
  1163, \dodoi{10.1086/498708}

\bibitem[{{Stalevski} {et~al.}(2012){Stalevski}, {Fritz}, {Baes}, {Nakos}, \&
  {Popovi{\'c}}}]{Stalevski_2012}
{Stalevski}, M., {Fritz}, J., {Baes}, M., {Nakos}, T., \& {Popovi{\'c}},
  L.~{\v{C}}. 2012, \mnras, 420, 2756, \dodoi{10.1111/j.1365-2966.2011.19775.x}

\bibitem[{{Stalevski} {et~al.}(2016){Stalevski}, {Ricci}, {Ueda}, {Lira},
  {Fritz}, \& {Baes}}]{Stalevski_2016}
{Stalevski}, M., {Ricci}, C., {Ueda}, Y., {et~al.} 2016, \mnras, 458, 2288,
  \dodoi{10.1093/mnras/stw444}

\bibitem[{{Taylor}(2005)}]{Taylor}
{Taylor}, M.~B. 2005, in Astronomical Society of the Pacific Conference Series,
  Vol. 347, Astronomical Data Analysis Software and Systems XIV, ed.
  P.~{Shopbell}, M.~{Britton}, \& R.~{Ebert}, 29

\bibitem[{{Treister} \& {Urry}(2006)}]{Treister2006}
{Treister}, E., \& {Urry}, C.~M. 2006, \apjl, 652, L79, \dodoi{10.1086/510237}

\bibitem[{{Voges} {et~al.}(1999){Voges}, {Aschenbach}, {Boller}, {Braeuninger},
  {Briel}, {Burkert}, {Dennerl}, {Englhauser}, {Gruber}, {Haberl}, {Hartner},
  {Hasinger}, {Kuerster}, {Pfeffermann}, {Pietsch}, {Predehl}, {Rosso},
  {Schmitt}, {Truemper}, \& {Zimmermann}}]{voges+99}
{Voges}, W., {Aschenbach}, B., {Boller}, T., {et~al.} 1999, VizieR Online Data
  Catalog, IX/10A

\bibitem[{{Voges} {et~al.}(2000){Voges}, {Aschenbach}, {Boller}, {Brauninger},
  {Briel}, {Burkert}, {Dennerl}, {Englhauser}, {Gruber}, {Haberl}, {Hartner},
  {Hasinger}, {Pfeffermann}, {Pietsch}, {Predehl}, {Schmitt}, {Trumper}, \&
  {Zimmermann}}]{voges+00}
---. 2000, VizieR Online Data Catalog, IX/29

\bibitem[{{Wada}(2012)}]{Wada_2012}
{Wada}, K. 2012, \apj, 758, 66, \dodoi{10.1088/0004-637X/758/1/66}

\bibitem[{{Wada} {et~al.}(2009){Wada}, {Papadopoulos}, \& {Spaans}}]{Wada_2009}
{Wada}, K., {Papadopoulos}, P.~P., \& {Spaans}, M. 2009, \apj, 702, 63,
  \dodoi{10.1088/0004-637X/702/1/63}

\bibitem[{{Watson} {et~al.}(2011){Watson}, {Denney}, {Vestergaard}, \&
  {Davis}}]{watson2011}
{Watson}, D., {Denney}, K.~D., {Vestergaard}, M., \& {Davis}, T.~M. 2011,
  \apjl, 740, L49, \dodoi{10.1088/2041-8205/740/2/L49}

\bibitem[{Wong {et~al.}(2019)Wong, Suyu, Chen, Rusu, Millon, Sluse, Bonvin,
  Fassnacht, Taubenberger, Auger, Birrer, Chan, Courbin, Hilbert, Tihhonova,
  Treu, Agnello, Ding, Jee, Komatsu, Shajib, Sonnenfeld, Blandford, Koopmans,
  Marshall, \& Meylan}]{10.1093/mnras/stz3094}
Wong, K.~C., Suyu, S.~H., Chen, G. C.-F., {et~al.} 2019, Monthly Notices of the
  Royal Astronomical Society, 498, 1420, \dodoi{10.1093/mnras/stz3094}

\bibitem[{{Wright} {et~al.}(2010){Wright}, {Eisenhardt}, {Mainzer}, {Ressler},
  {Cutri}, {Jarrett}, {Kirkpatrick}, {Padgett}, {McMillan}, {Skrutskie},
  {Stanford}, {Cohen}, {Walker}, {Mather}, {Leisawitz}, {Gautier}, {McLean},
  {Benford}, {Lonsdale}, {Blain}, {Mendez}, {Irace}, {Duval}, {Liu}, {Royer},
  {Heinrichsen}, {Howard}, {Shannon}, {Kendall}, {Walsh}, {Larsen}, {Cardon},
  {Schick}, {Schwalm}, {Abid}, {Fabinsky}, {Naes}, \& {Tsai}}]{wright+10}
{Wright}, E.~L., {Eisenhardt}, P. R.~M., {Mainzer}, A.~K., {et~al.} 2010, \aj,
  140, 1868, \dodoi{10.1088/0004-6256/140/6/1868}

\end{thebibliography}
\bibliographystyle{aasjournal}


\appendix
\counterwithin{figure}{section}
\section{Cumulative probability distribution}
\begin{figure*}[hbt!]
    \centering
    \includegraphics[scale=0.26]{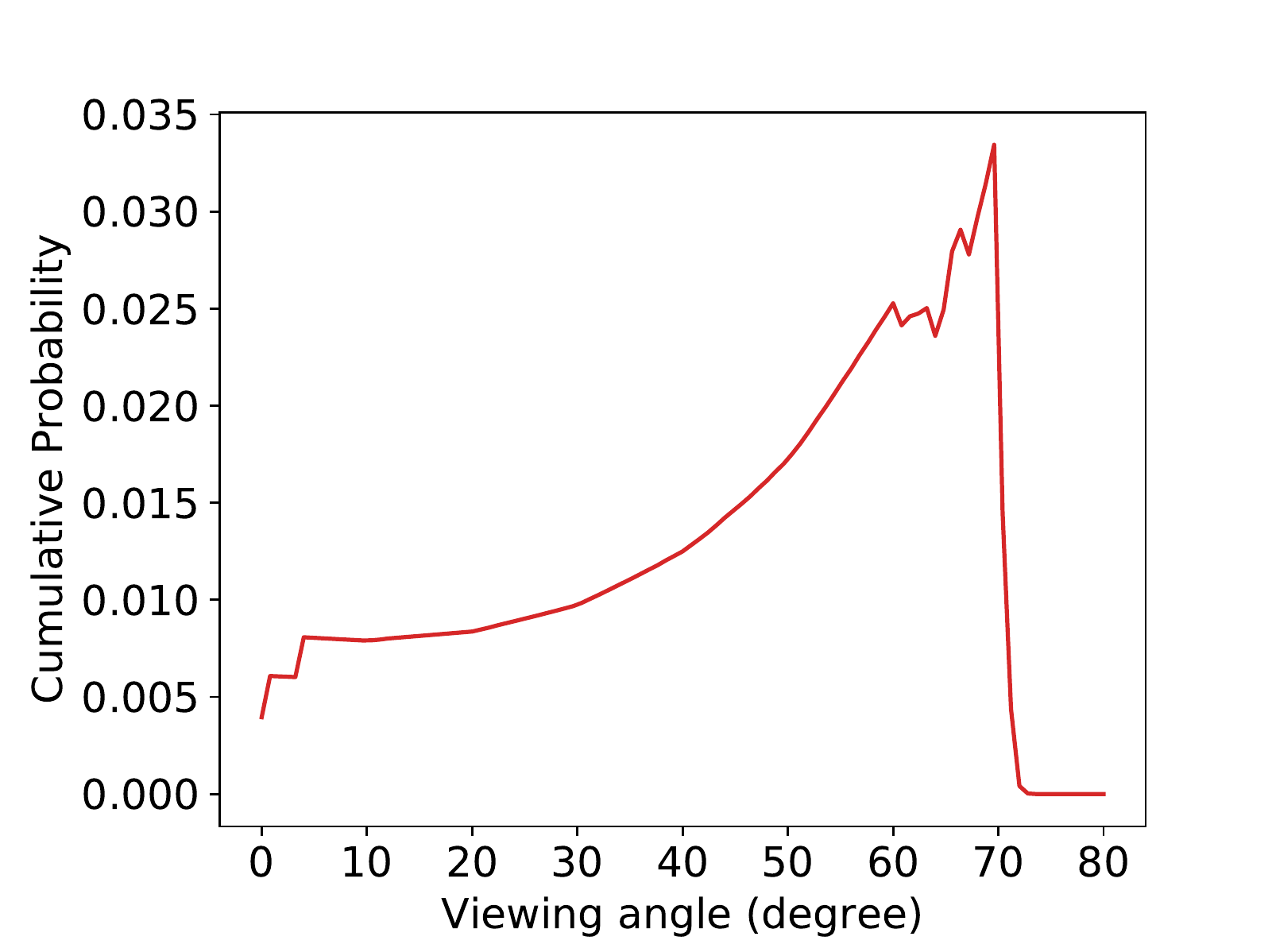}
    \includegraphics[scale=0.26]{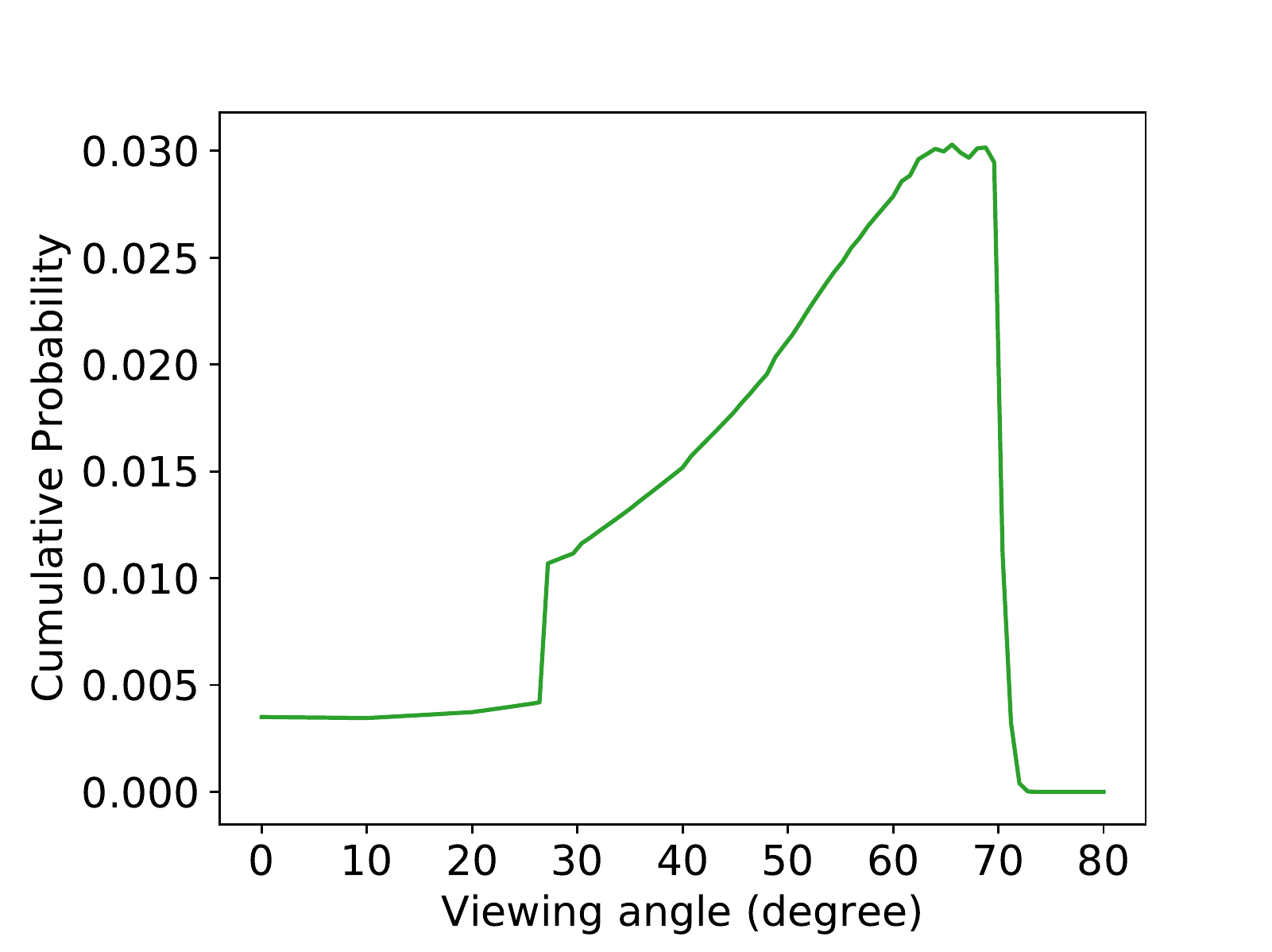}
    \includegraphics[scale=0.26]{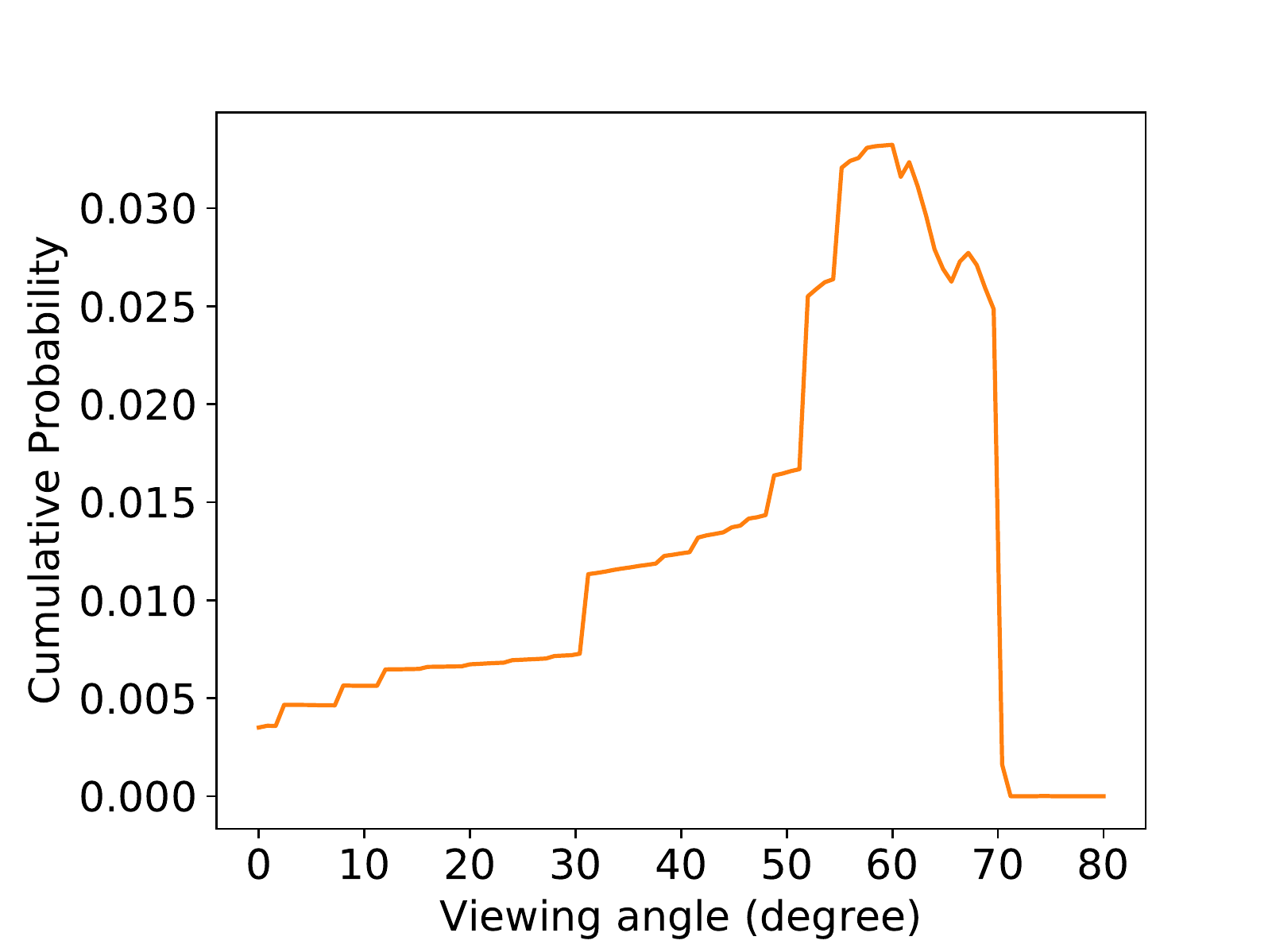}
    \includegraphics[scale=0.26]{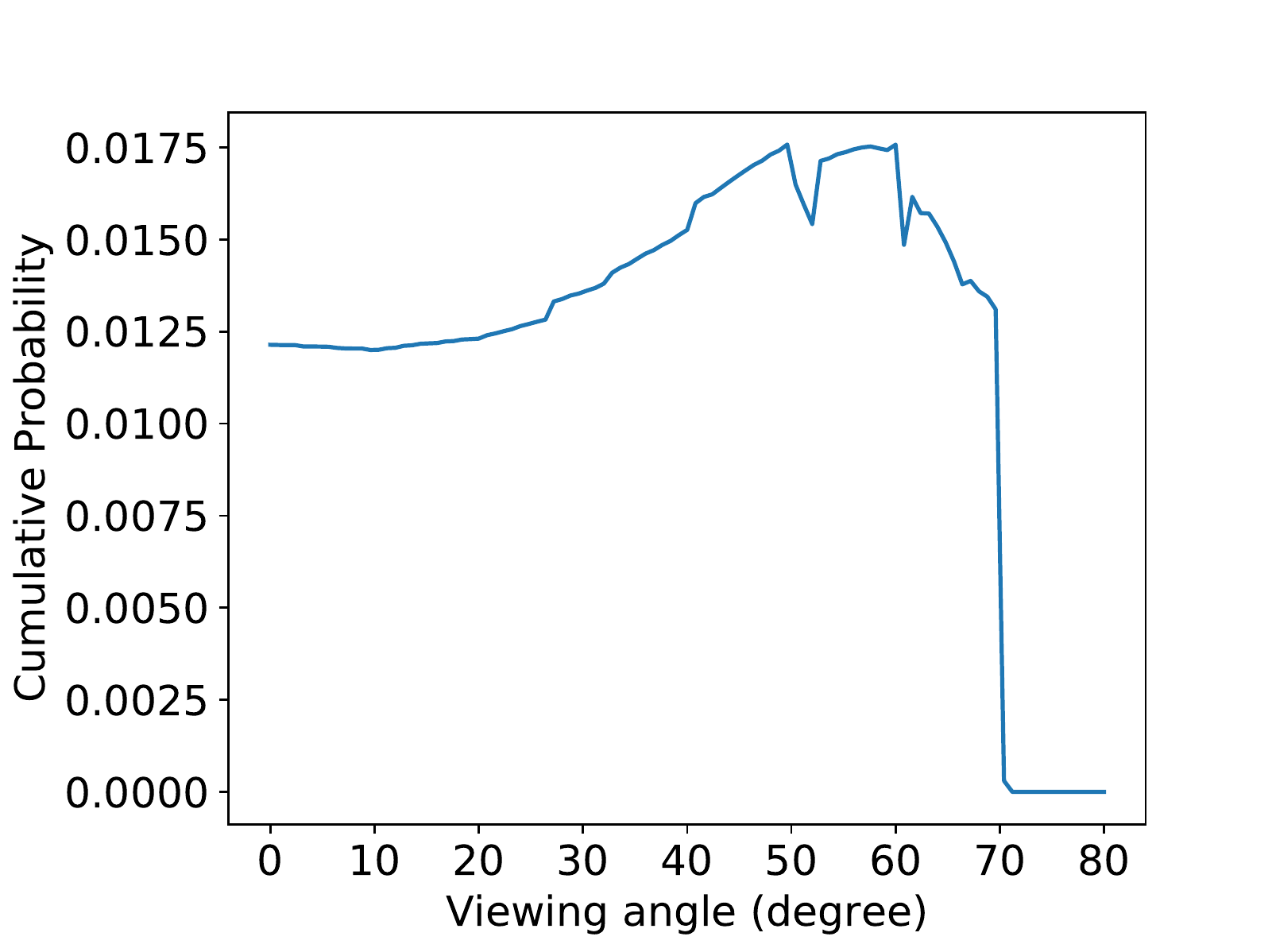}
    \includegraphics[scale=0.26]{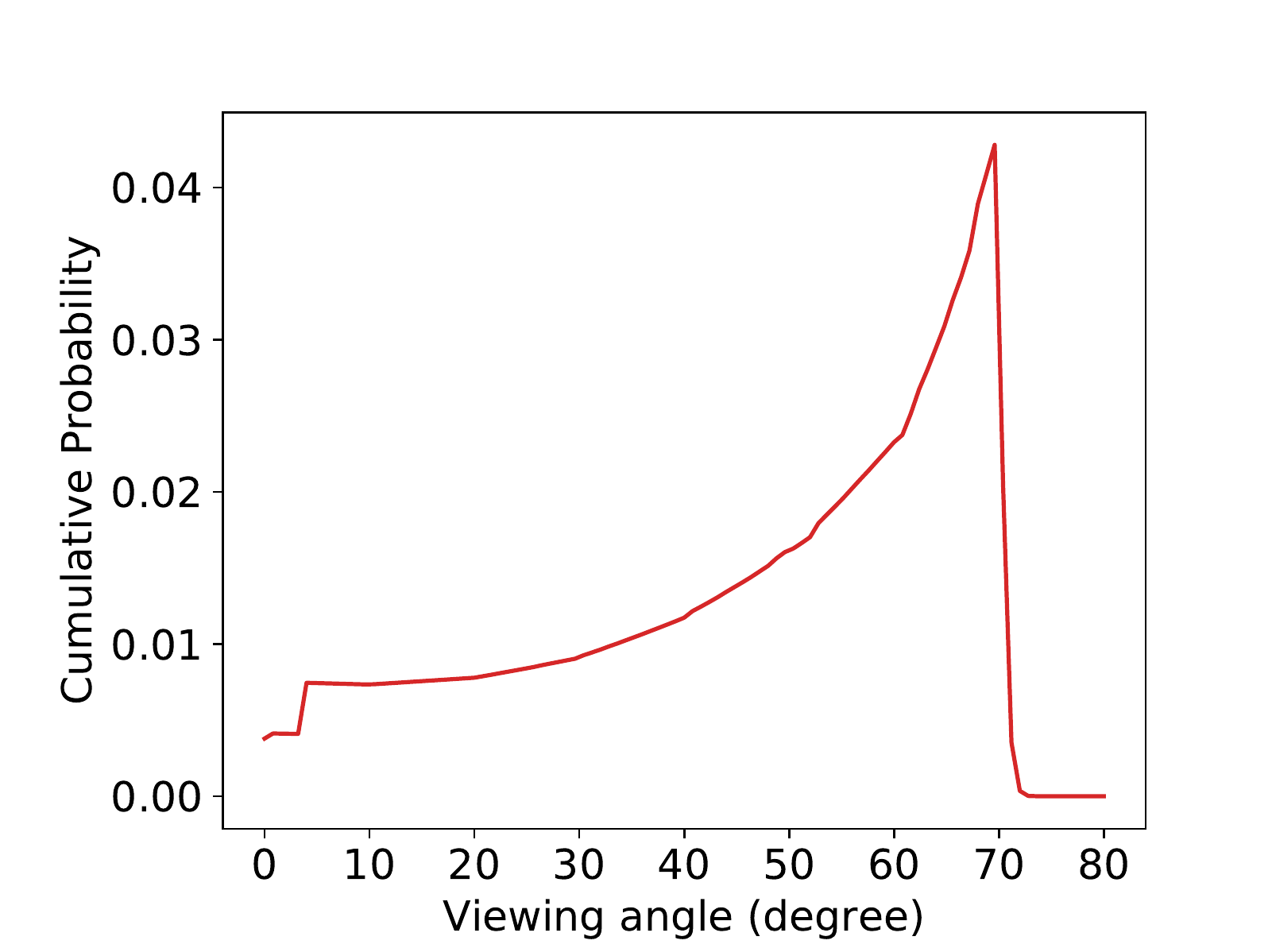}
    \includegraphics[scale=0.26]{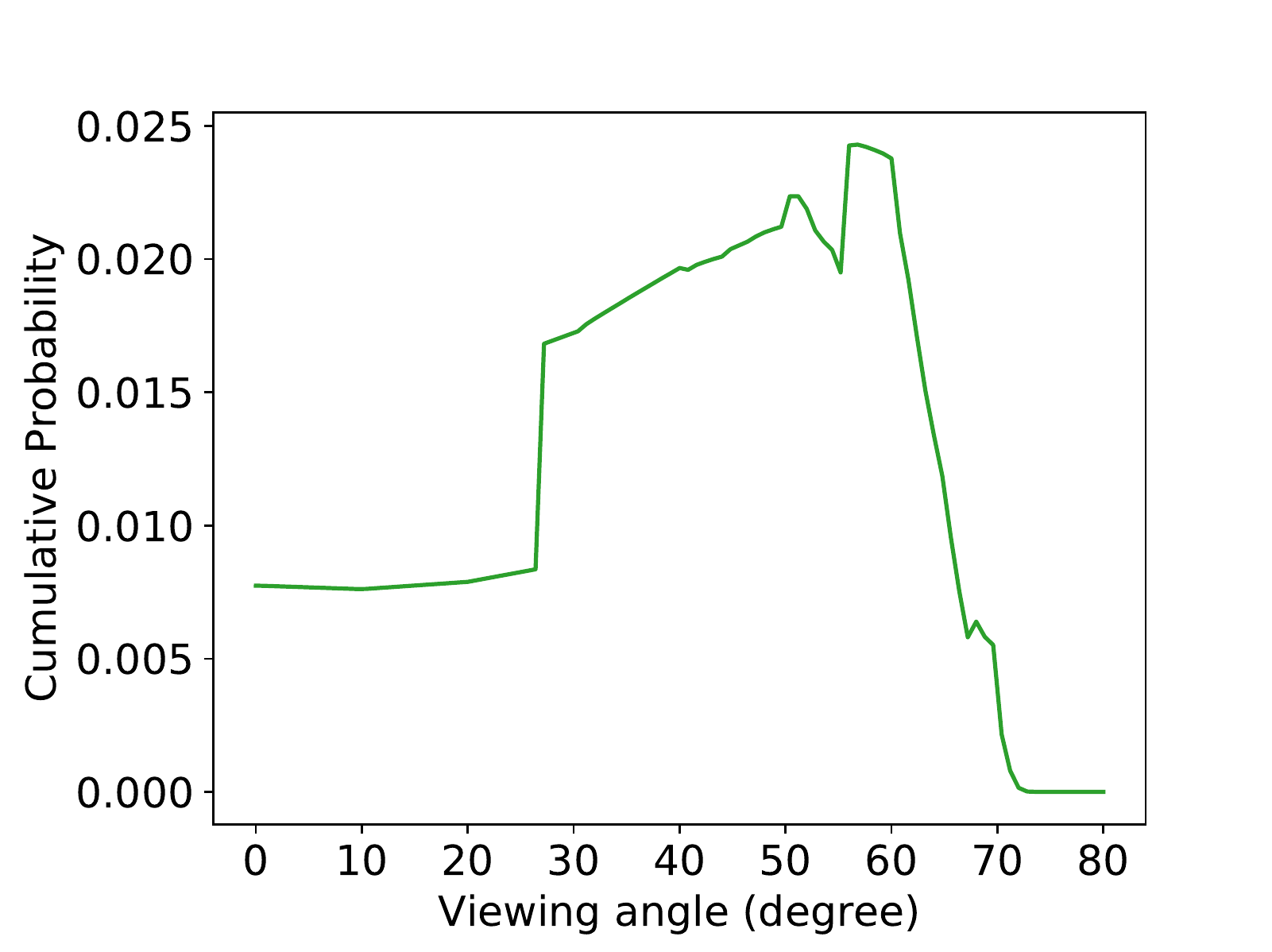}
    \includegraphics[scale=0.26]{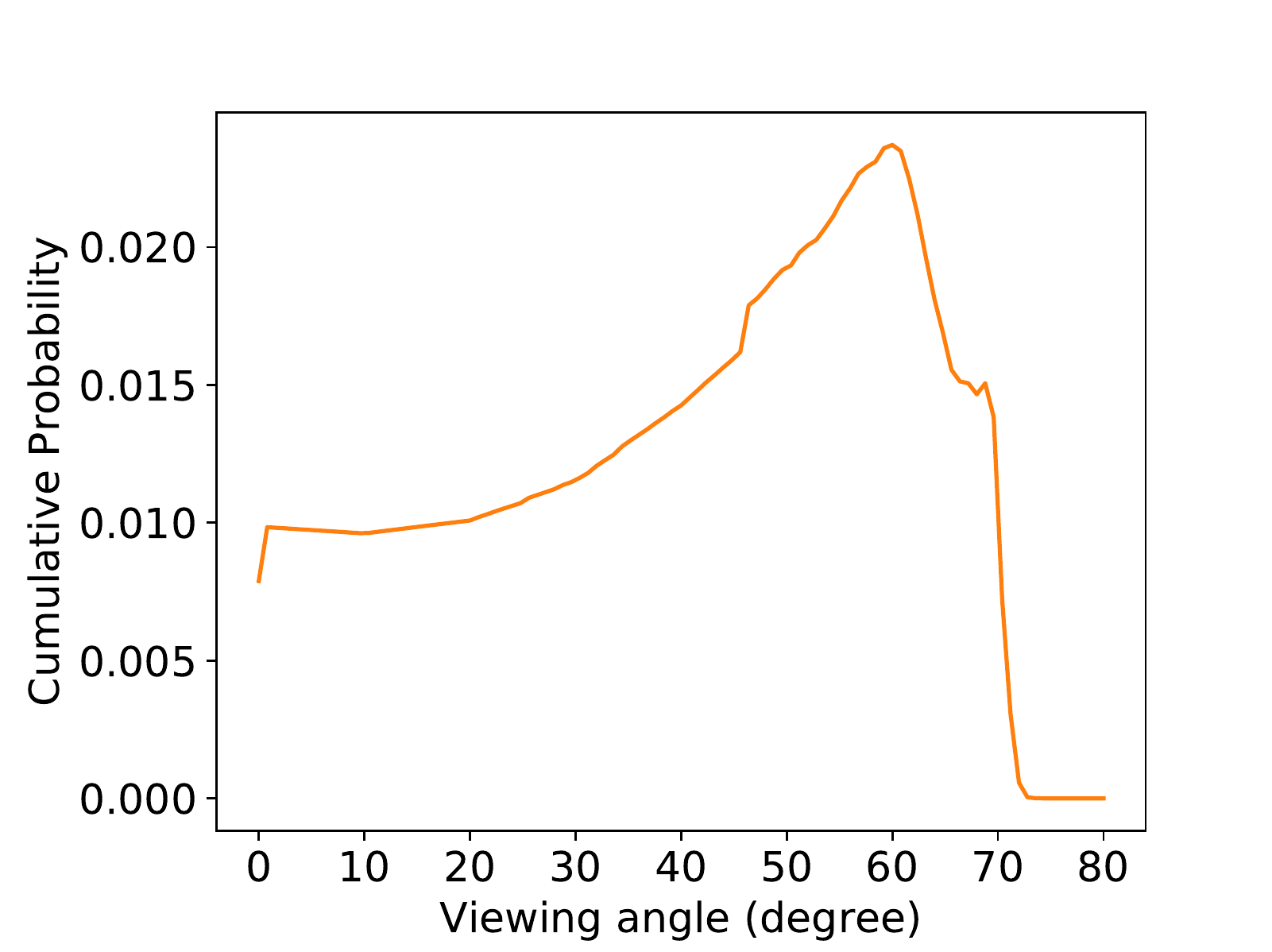}
    \includegraphics[scale=0.26]{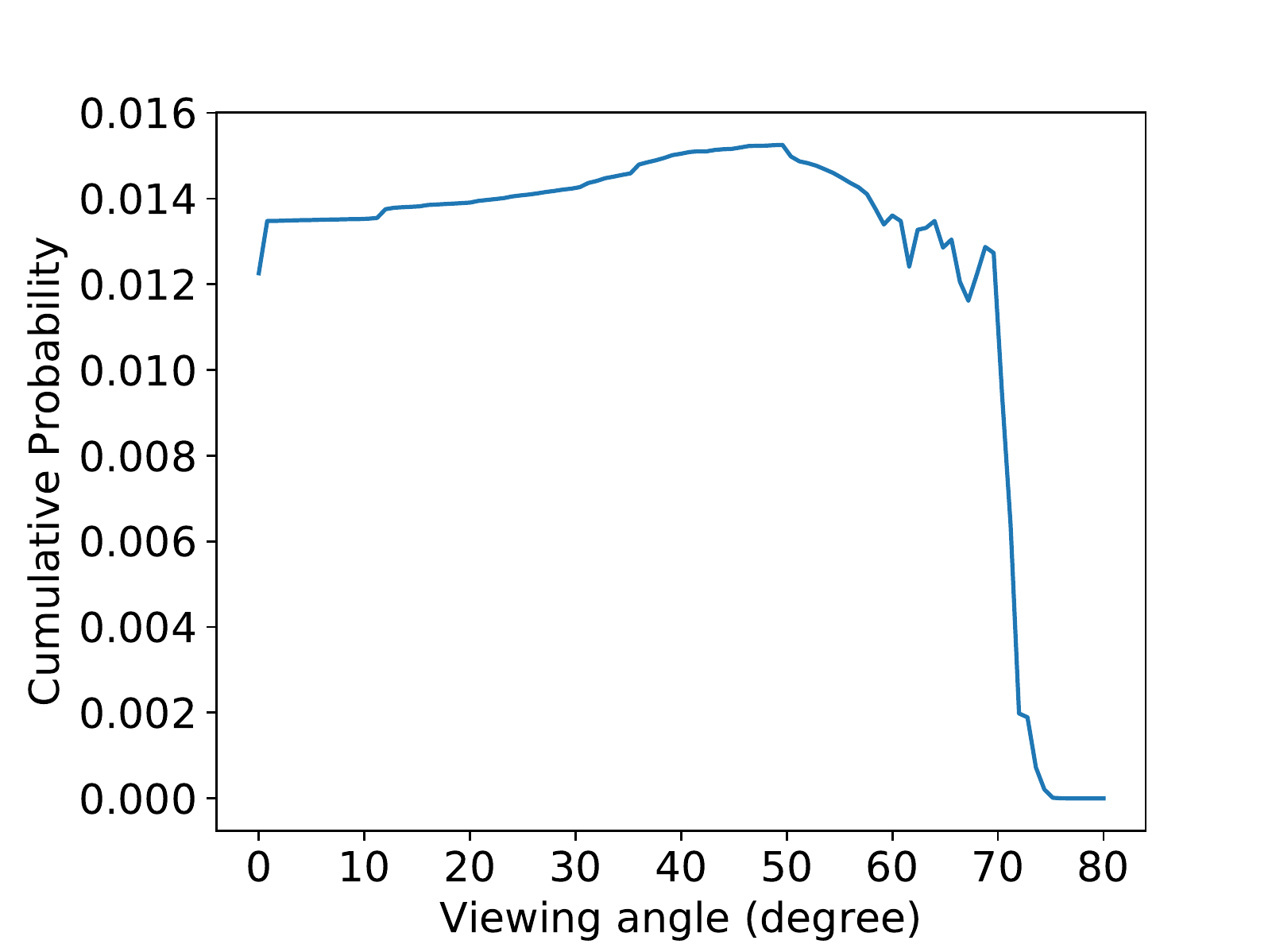}
    \includegraphics[scale=0.26]{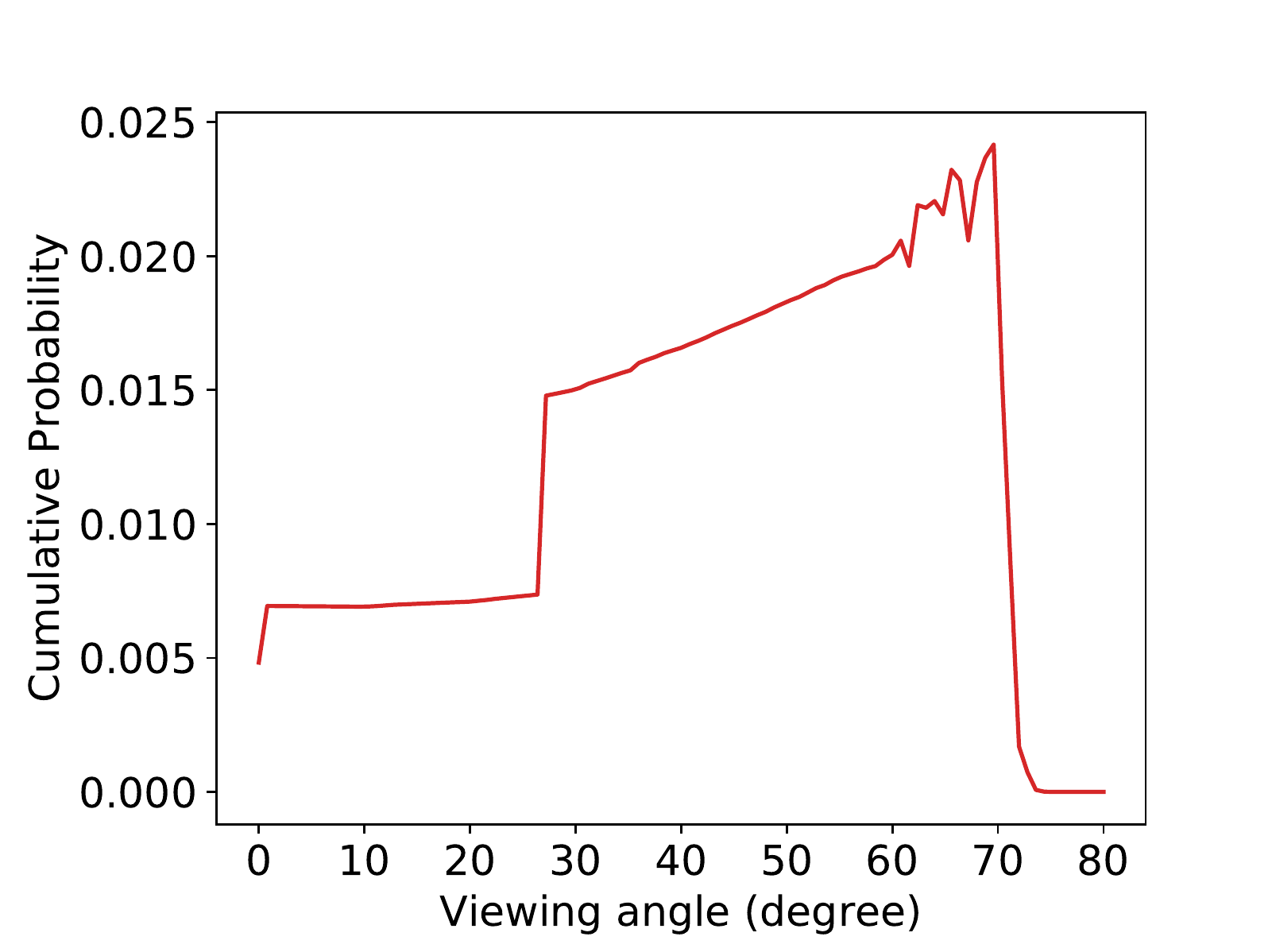}
    \includegraphics[scale=0.26]{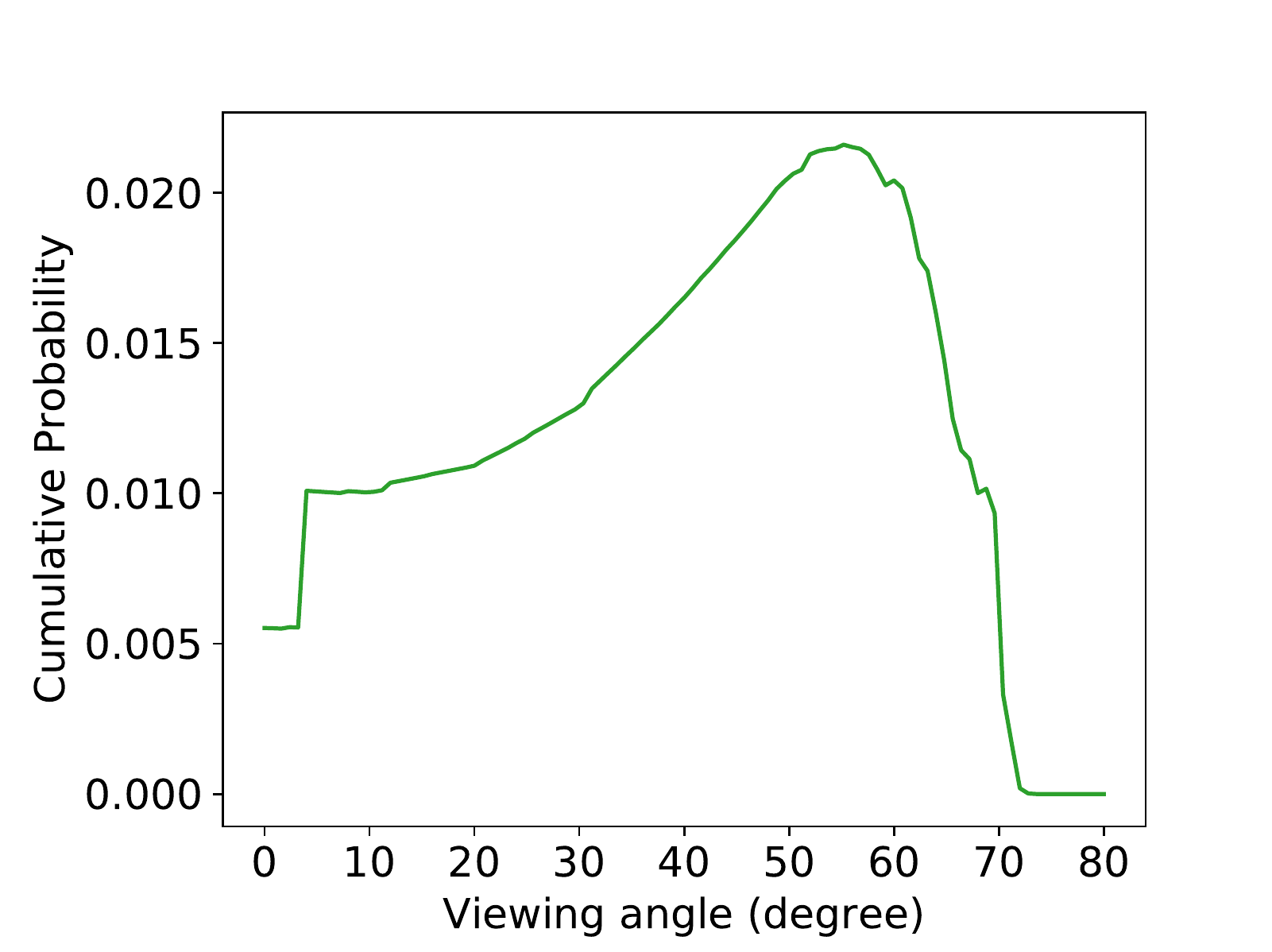}
    \includegraphics[scale=0.26]{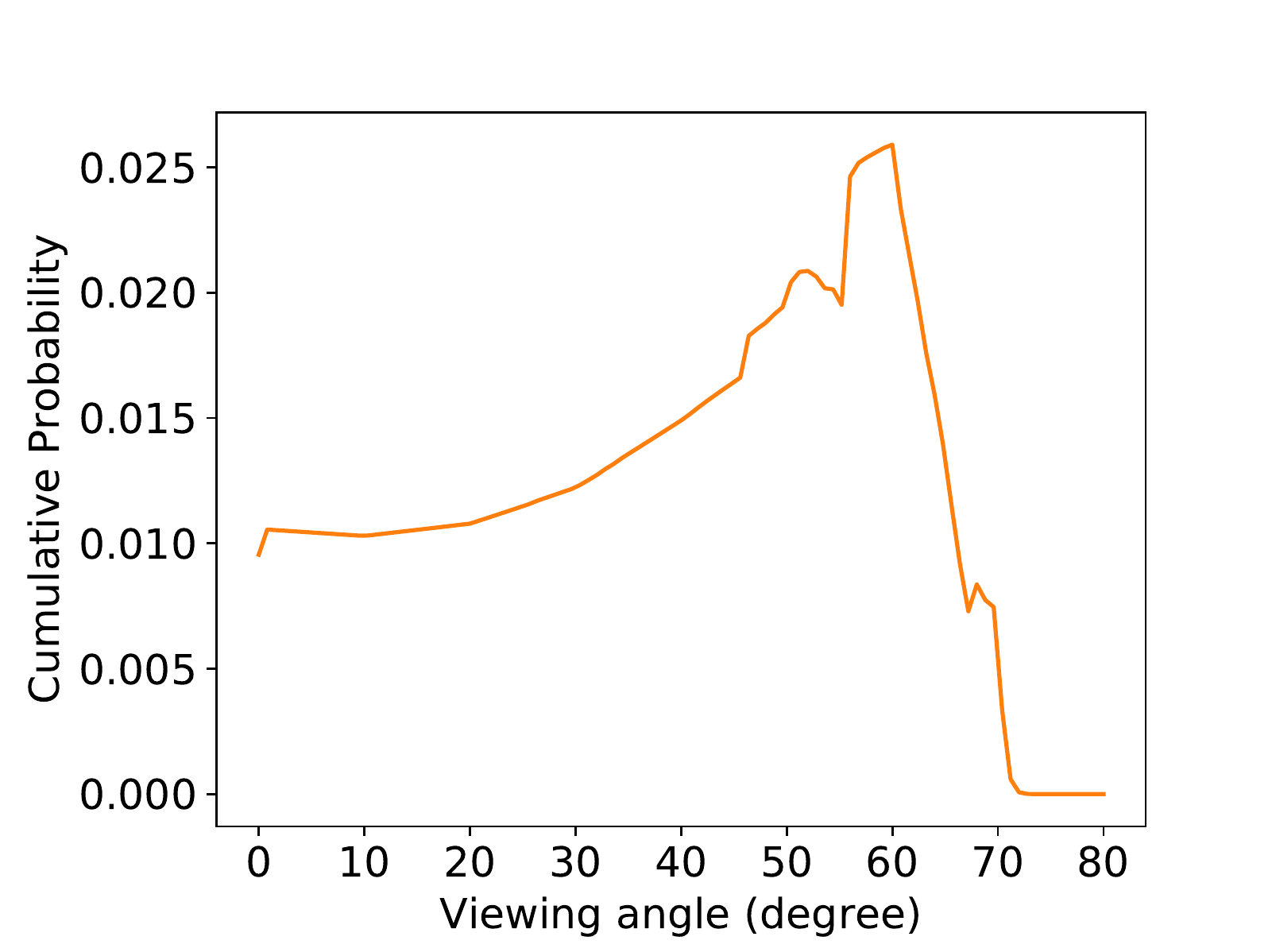}
    \includegraphics[scale=0.26]{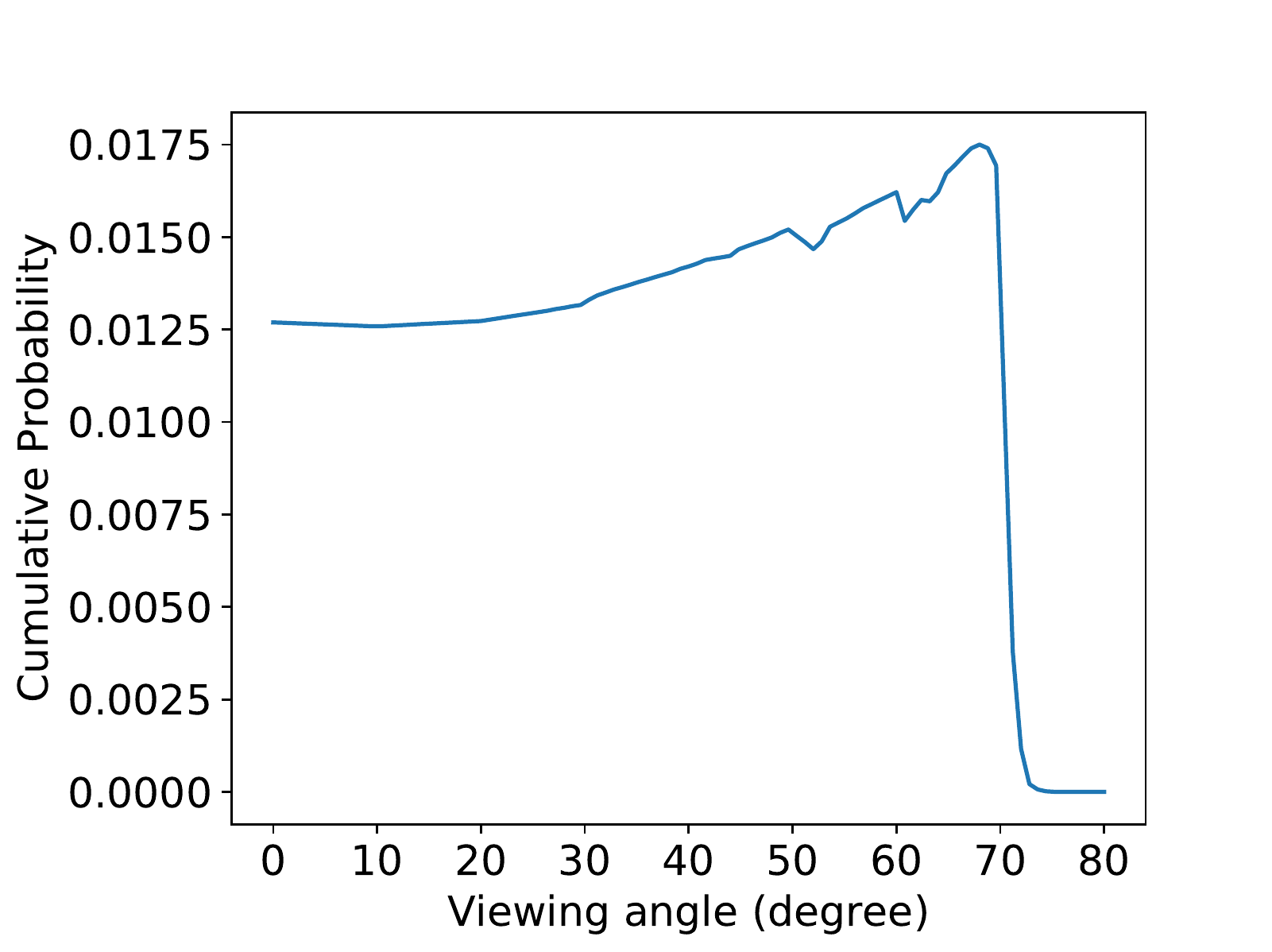}
    \caption{Normalized cumulative probability distribution of redshift (upper panel), black hole mass (middle panel), and Eddington ratio (lower panel) for each bins. Column 1$-$4 represents the bin1$-$bin4. The y-axis in the plot is cumulative probability and the x-axis is the viewing angle in degrees.}
    \label{fig:comulative_prob}
\end{figure*}

Here, we present the detail calculation of the viewing angle in each bins of redshift, mass, and Eddington ratio. The figure\ref{fig:comulative_prob} shows all the cumulative probability distribution that we have used to estimate the viewing angle.

\begin{figure}
    \centering
    \includegraphics[scale=0.4]{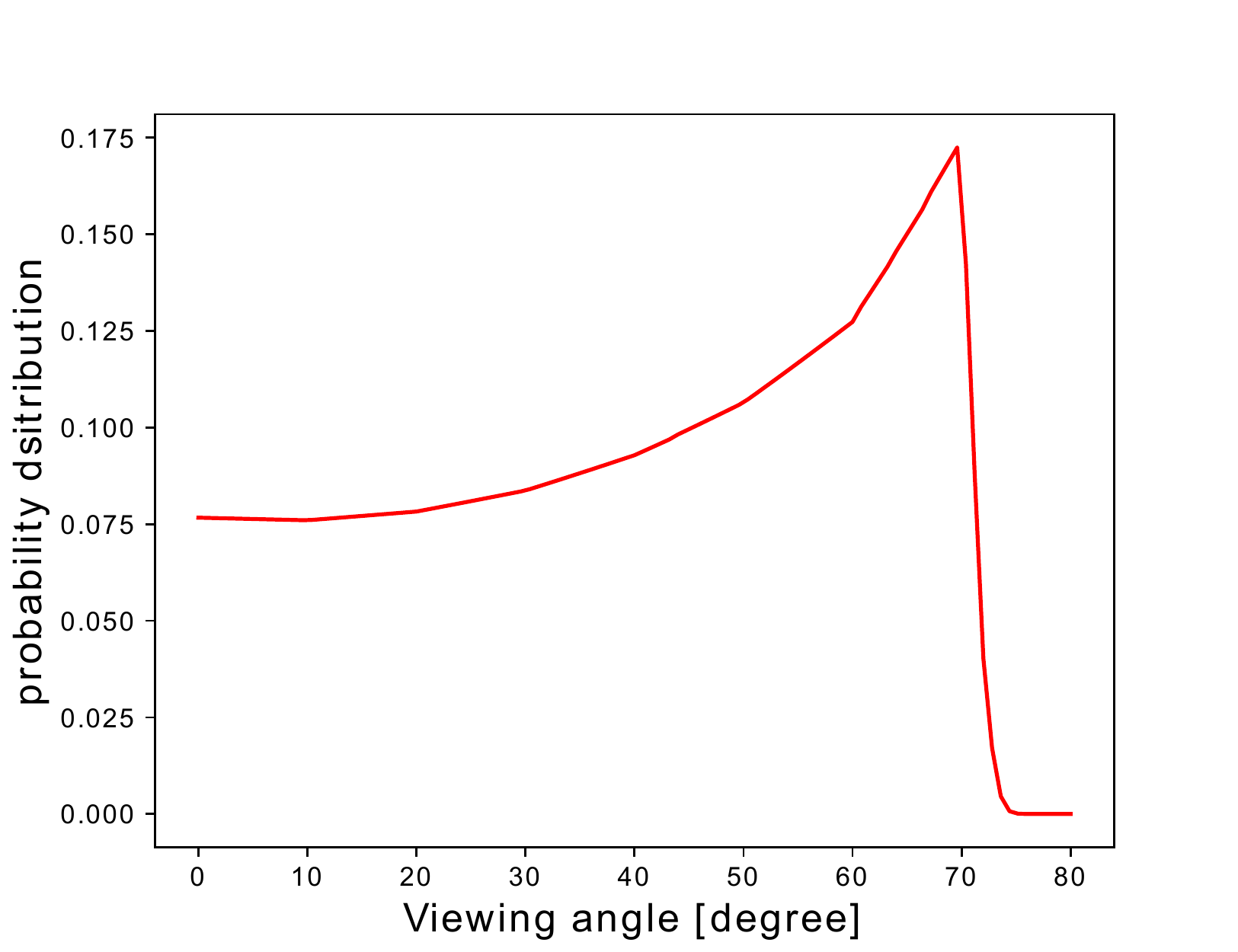}
    \caption{Normalized probability distribution of an individual source, J095821.65+024628.1. The y-axis in the plot is probability and the x-axis is the viewing angle in degrees. }
    \label{fig:normprob}
\end{figure}
\end{document}